\documentclass[10,journal]{IEEEtran}
\usepackage{amsmath,amsfonts}
\usepackage{algorithmic}
\usepackage{algorithm}
\usepackage{array}
\usepackage{textcomp}
\usepackage{url}
\usepackage{verbatim}
\usepackage{graphicx}
\usepackage{cite}

\usepackage{caption}
\usepackage{subcaption}
\usepackage{booktabs}
\usepackage[table,xcdraw]{xcolor}
\usepackage{multirow}
\usepackage{makecell}
\usepackage{enumitem}
\usepackage{bm}
\usepackage{amssymb}
\usepackage{amsthm}

\usepackage{soul}

\theoremstyle{plain}
\newtheorem{thm}{Theorem}
\newtheorem{lem}[thm]{Lemma}

\newenvironment{NewProof}{{\noindent\it Proof.}}{\hfill $\blacksquare$\par}
\allowdisplaybreaks[4]

\ifodd 1
\else
\fi

\begin{document}

\title{Optical Integrated Sensing and Communication \\ with Light-Emitting Diode}

\author{
Runxin Zhang,
Yulin~Shao,
Menghan Li,
Lu Lu,
Yonina C. Eldar
\thanks{R. Zhang, M. Li, and L. Lu are with the Key Laboratory of Space Utilization, Technology and Engineering Center for Space Utilization, Chinese Academy of Sciences, Beijing, 100094, China, and with the University of Chinese Academy of Sciences, Beijing, 100049, China (e-mail: zhangrunxin20@mails.ucas.ac.cn).

Y. Shao is with the State Key Laboratory of Internet of Things for Smart City, University of Macau, Macau S.A.R, China, 
and with the Department of Electrical and Electronic Engineering, Imperial College London, London SW7 2AZ, U.K. (e-mail: ylshao@um.edu.mo).

Y. C. Eldar is with the Weizmann Institute of Science, Rehovot 7610001, Israel (e-mail:yonina.eldar@weizmann.ac.il).
}
}



\maketitle

\begin{abstract}
This paper presents a new optical integrated sensing and communication (O-ISAC) framework tailored for cost-effective Light-Emitting Diode (LED) for enhanced Internet of Things (IoT) applications. Unlike prior research on ISAC, which predominantly focused on radio frequency (RF) band, O-ISAC capitalizes on the inherent advantages of the optical spectrum, including the ultra-wide license-free bandwidth, immunity to RF interference, and energy efficiency -- attributes crucial for IoT communications. The communication and sensing in our O-ISAC system unfold in two phases: directionless O-ISAC and directional O-ISAC. In the first phase, distributed optical access points emit non-directional light for communication and leverage small-aperture imaging principles for sensing. In the second phase, we put forth the concept of optical beamforming, using collimating lenses to concentrate light, resulting in substantial performance enhancements in both communication and sensing. Numerical and simulation results demonstrate the feasibility and impressive performance of O-ISAC benchmarked against optical separate communication and sensing systems.
\end{abstract}


\section{Introduction}\label{sec:I}

In today's increasingly interconnected world, the integration of communication and sensing systems into the fabric of Internet of Things (IoT) ecosystems has become crucial for the development of smart environments \cite{zhang20196g,strinati20196g,denoising}. As we delve deeper into the era of IoT, the seamless and intelligent interaction between devices and their environments necessitates advancements in technologies that can simultaneously support communication and sensing capabilities.

In this context, one technique that has garnered significant attention is integrated sensing and communication (ISAC) \cite{liu2022survey,cui2021integrating,wang2022towards}. The core idea behind ISAC involves utilizing the time and frequency resources originally allocated to radar, and developing a unified transceiver that achieves wireless communication and remote sensing simultaneously with a single hardware platform \cite{liu2020joint, cui2021integrating, wang2022towards}. While most prior research in ISAC focused on radio frequency (RF) ISAC \cite{xiao2023integrated,wu2022integrating,zhang2022integrated,bazzi2023integrated,gunlu2023secure}, recent papers have begun to explore the integration of communication and sensing within a much higher frequency band -- the optical band\cite{wen2024optical,liang2023integrated,yan2023technical}.

Compared to radio communication, the optical band is ultra-wide and license-free \cite{rahman2020review} -- it consists of three sub-bands: the infrared (IR), the visible, and the ultraviolet (UV). Therefore, optical communication is viewed as a promising complement to radio communication in next-generation communication systems \cite{chow2024recent} to address the problem of spectrum scarcity. 
Recent advancements in light-emitting diode (LED)-based Optical Wireless Communication (OWC), notably Visible Light Communication (VLC), have catalyzed significant developments that are particularly advantageous for IoT applications.  
On the other hand, optical sensing is an estimation technology that measures the presence of objects, distance, displacement, etc. \cite{pathak2015visible}, with high precision by capturing the intensity of incident light rays and converting it into a form readable by a measuring device.
Particularly, given the limitations of GPS in indoor IoT scenarios, the optical spectrum's wide bandwidth and electromagnetic interference immunity offer significant advantages for visible light positioning \cite{huang2022indoor}.

Since optical communication and optical sensing both rely on the same fundamental medium -- light, there is growing recognition in research of the potential synergies and efficiencies that could be achieved by integrating these two essential functions, including both wired optical integrated communication and sensing (O-ISAC), e.g., fiber-optic ISAC \cite{yan2023technical,chen2022photonic,he2023integrated}, and wireless O-ISAC \cite{cao20234,cao2022unified,lyu2023radar}.
The design of wireless O-ISAC systems primarily focuses on laser-based light sources,
adhering to the principles of RF-ISAC.
Ref. \cite{cao2022unified}, for example, proposed a unified waveform based on quadrature phase shift keying (QPSK) and direct sequence spread spectrum for joint waveform modulation.
Additionally, orthogonal electromagnetic polarizations \cite{lei2023integration} and time-division multiplexing \cite{li2024sensing} are employed to enable the sensing and communication functions to share the same transceiver hardware.
It is important to note that the primary focus of these works is on laser-based O-ISAC systems, which do bear similarities to RF systems, resulting in a more straightforward design process. 

Considering the hardware costs and wide availability of LED, \cite{shi2022joint} proposes an LED-based ISAC system architecture. This system integrates optical communication with positioning using multi-band carrierless amplitude and phase (m-CAP) modulation and received signal strength (RSS)-based trilateration.
Similarly, \cite{zhang2023optical} introduces a converged underwater wireless system that combines optical communication with sensor networks. Target positioning in this system is determined by calculating cross-correlation values (CCV) based on RSS.

While these LED-based systems effectively combine optical wireless communication and optical sensing, their integration is presently limited to concurrent use of transmitted signals. However, we argue that O-ISAC systems extend beyond mere hardware and signal sharing. These systems are inherently complementary: optical communication provides essential illumination for optical sensing, and optical sensing in turn supplies environmental information that can significantly enhance optical communication. This dual functionality not only maximizes resource utilization but also opens up new avenues for innovative applications in diverse environments.


{\it Contributions:} This paper develops a new O-ISAC framework tailored for cost-effective commercial LED. Our driving force is to unlock the untapped potential found at the intersection of optical communication and optical sensing, transcending the conventional boundaries and paving the way for more intelligent and resource-efficient O-ISAC systems.
Compared to RF-ISAC and O-ISAC with laser, we point out several notable challenges of O-ISAC using LED as following:
\begin{itemize}[leftmargin=0.6cm]
\item {\it Incoherent light}. LEDs emit incoherent light, rendering phase modulation and coherent detection unattainable.  Additionally, this incoherence makes beamforming impossible, as we cannot manipulate the phase of light emitted by distinct LEDs.

\item {\it Divergent light}. LEDs emit light in a scattered manner, covering a broad area. This divergence leads to a significant reduction in the intensity of light received by the target device, even when the LED is aligned directly to the target.

\item {\it Massive echo}. Unlike conventional RF-ISAC systems that assume a limited number of scatters (echoes), we consider a more realistic setup, where all objects in the environment can be reflectors and the reflected light from all range bins of the environment can be collected by an optical sensor.

\item {\it Frequency dispersion}. Commercial LEDs are unable to produce monochromatic light signals, a limitation that presents a considerable obstacle in managing the full spectrum of emitted light.
\end{itemize}

In addressing the above unique challenges and unlocking the full potential of O-ISAC, we explore pragmatic strategies that address each of these specific hurdles.
Our main contributions are summarized as follows.

We put forth a new O-ISAC framework tailored for cost-effective commercial LEDs. 
To tackle the challenge of incoherent light, we leverage the non-coherent detection method, intensity modulation and direct detection (IM/DD), for optical communication, and the pinhole imaging principle, which relies solely on light intensity detection, for optical sensing.
Another notable advantage of the pinhole imaging sensing principle lies in its ability to project reflected light from both the object of interest and unwanted objects onto distinct spatial coordinates within a pinhole plane. This spatial isolation of the echo signals reduces interference from non-target scatters, thereby addressing the massive echo challenge.

Within the O-ISAC framework, we reveal the synergistic effect of optical communication and optical sensing -- they share the common goal of maximizing the received light intensity at the target device.
Recognizing this synergy, we formulate a received light intensity maximization problem to address the challenge of divergent light emission from LEDs.
To solve the maximization problem, our proposed approach involves a meticulous optimization of both the source layout and the radiation pattern of LEDs.

\begin{itemize}[leftmargin=0.6cm]
    \item {\it Source layout optimization}. In contrast to conventional source layout optimization in optical communication, we propose a new optimization criterion that maximizes the coverage area where the received light intensity exceeds a threshold. Given the new criterion, we analytically solve the optimization problem and give a closed-form approximation to the optimal source distribution.
    \item {\it LED radiation pattern optimization}.  To achieve spatial selectivity, we develop the concept of optical beamforming by means of collimating lenses, a class of curved optical lenses that can align light rays in a parallel fashion. With optical beamforming, directionless light emitted by LEDs is steered towards the target device, offering a substantial enhancement in the received light intensity. To derive the optimal radiation pattern after optical beamforming, we address the challenge of the frequency dispersion effect of LEDs, and analytically characterize the optimal profile of the lens surface and the angle of departure (AoD) of different frequency components in closed forms.
\end{itemize}

Overall, these two optimization problems give rise to a two-phase operational mechanism, which essentially shapes the ultimate design of the proposed O-ISAC framework.
In phase 1, the system utilizes directionless light rays to broadcast control signals to all devices within the coverage, while multiple distributed sensors collect reflected light and estimate the device positions. The LED locations are optimized following the solution of source layout optimization.
In the second phase, a refined operation unfolds, and each device is individually served in a Time Division Multiple Access (TDMA) fashion. The light intensities are significantly enhanced by optical beamforming, leading to more reliable communication and much more accurate device sensing and tracking.

Numerical and simulation results confirm the superior performance of the proposed O-ISAC framework. 
Compared with a separate communication and sensing system, the source layout optimization in the first phase yields a remarkable improvement in both communication bit error rate (BER) and sensing mean squared error (MSE) -- the gains are up to $3.47$ dB and $3.14$ dB, respectively. 
In the second phase, the introduction of optical beamforming further optimizes the radiation patterns of LEDs, yielding substantial performance gains over the separate communication and sensing system. The light intensity falling on the target object is increased by $65.45\%$, the communication BER is improved by $63.35$ dB, and the sensing MSE is improved by $40.42$ dB. This significant gain is entirely consistent with our expectations, thanks to the greatly concentrated light facilitated by optical beamforming.

{\it Organization}: The remainder of this paper is structured as follows.
Section~\ref{sec:II} overviews the system model and formulates the processes of optical communication and optical sensing.
Section~\ref{sec:III} reveals the synergies between optical sensing and optical communication, and proposes the two-phase operation mechanism.
The source layout optimization and radiation pattern optimization are detailed in Sections~\ref{sec:III} and \ref{sec:IV}, respectively.
Section~\ref{sec:V} presents the numerical and simulation results.
Section~\ref{sec:VI} concludes this paper.

{\it Notations}: We use boldface lowercase letters to denote column vectors (e.g., $\bm{x}$, $\bm{v}$) and boldface uppercase letters are matrices (e.g., $\bm{X}$, $\bm{S}$).
For a vector or matrix, 
$(\cdot)^\top$ is the transpose, 
and $(\cdot)^H$ is the conjugate transpose.
$\mathbb{R}$ and $\mathbb{C}$ stand for the sets of real and complex values, respectively,
$c$ represents the speed of light, and $m_0$ is Lambert’s mode number.
The vectorization function and de-vectorization function are written as $\emph{vec}(\cdot)$ and $\emph{devec}(\cdot)$,
$(\cdot) \otimes (\cdot)$ denotes the Kronecker product, 
and $\delta(\cdot)$ is the Dirac delta function.

\section{Problem Formulation}\label{sec:II}
\begin{figure}
    \centering
    \includegraphics[width=0.8\linewidth]{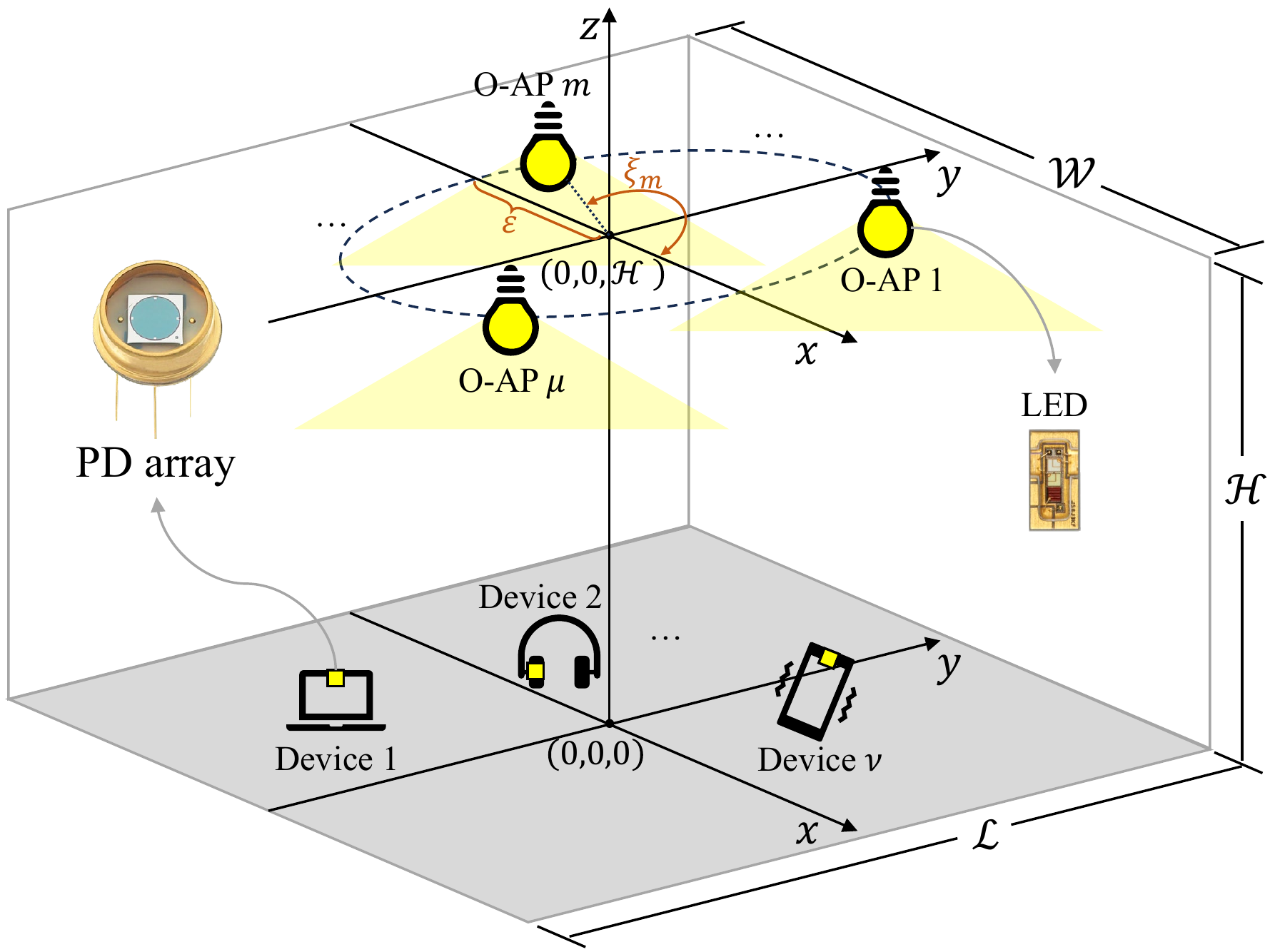}	
    \caption{The system model of the proposed O-ISAC framework. 
}
    \label{f:model}
\end{figure}

We consider an indoor IoT scenario with $\mu$ distributed optical access points (O-APs), where each O-AP is equipped with an optical source and an optical sensor. In our O-ISAC framework, we designate the optical source as an LED and the optical sensor as a pinhole camera, as shown in Fig.~\ref{f:model}.
The dimension of the room is $\mathcal{W}\times\mathcal{L}\times \mathcal{H}$.
The O-APs are arranged in a circular pattern,
with the ceiling's center serving as the center and a radius of $\varepsilon$.
Therefore, the coordinates of the O-APs (hence the LEDs and pinhole cameras) are 
\begin{equation}
    \bm{p}_{\text{OAP},m} \triangleq (\varepsilon \cos{\xi_m},\allowbreak \varepsilon \sin{\xi_m},\allowbreak \mathcal{H}),~m=1,2,..., \mu,
\end{equation}
where $\xi_m$ denotes the angle between the $m$-th O-AP and the positive x-axis, as shown in Fig.~\ref{f:model}.
Let there be $\nu$ devices in the room, and each device is equipped with a square photodiode (PD) array containing $\kappa$ PDs as the optical receiver.

Our O-ISAC framework harnesses the emitted light from LEDs as the carrier of information, enabling downlink data transmission to devices. The light reflected by these devices is then captured by pinhole cameras, providing us with the means to perceive and comprehend device states. The utilization of light as the communication and sensing medium introduces a distinct array of challenges in system design, setting it apart from the conventional RF-ISAC. Nonetheless, it also brings forth numerous new opportunities to circumvent the issues faced by RF-ISAC.
In the following, we will formulate the optical communication and sensing processes, respectively, in more detail and explore its potential of O-ISAC.

\begin{figure}
    \centering
    \includegraphics[width=0.45\linewidth]{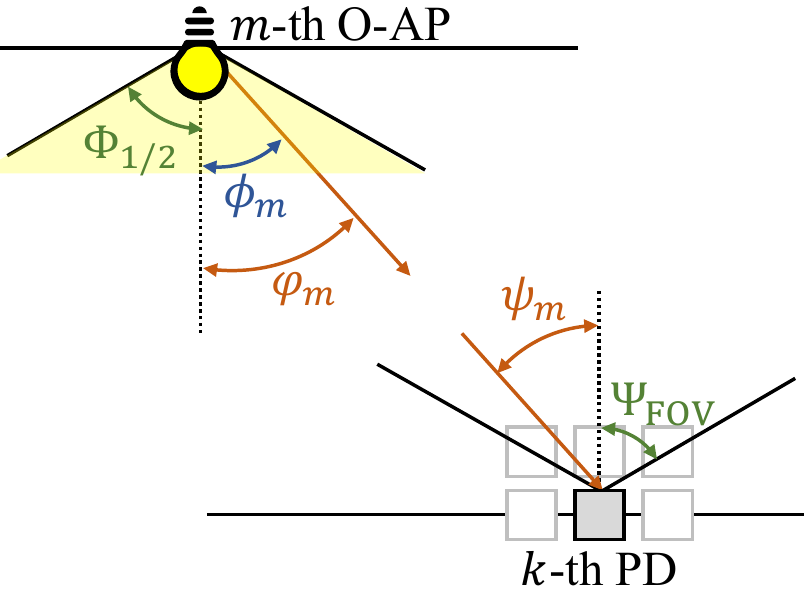}
    \caption{Illustrating the propagation model of an LED source. 
}
    \label{f:angle}
\end{figure}

\subsection{Propagation model of light source}
The radiation patterns of LED light sources are starkly different from those of RF antennas. Therefore, our first step involves modeling the propagation patterns of LED to effectively analyze the received signal strength at the devices.

We adopt the Lambertian model \cite{djordjevic2018advanced} to characterize the radiation pattern of LED.
Consider the $m$-th O-AP. As depicted in Fig.~\ref{f:angle}, the LED has a semi-angle at half power, denoted by $ \Phi_{1/2}$, and the angle of emission (AoE) of light beams $\phi_{m}$ is constrained within $0 \le \phi_{m}\le \Phi_{1/2}$.
The angle of departure (AoD) of a light beam is denoted by $\varphi_m$, and from Fig.~\ref{f:angle} we have $\phi_m=\varphi_m$.
Later in Section \ref{sec:IV}, we will integrate collimating lenses into our setup to manipulate the radiation pattern of the emitted light, in which case the direction of the light beam can be steered, and $\varphi_m$ is different from $\phi_m$.

Consider a specific device. We index the $\kappa$ PDs by $\{k:k=1,2,...,\kappa\}$ and denote the coordinates of these PDs by $\bm{p}_{P,k} = (x_{P,k} , y_{P,k} , z_{P,k})$. 
As depicted in Fig.~\ref{f:angle}, each PD has a half-angle field-of-view (FOV) denoted by $\Psi_{\text{FOV}}$, and the angle of arrival (AoA) $\psi_m$: $0 \le \psi_{m}\le \Psi_{\text{FOV}}$.
When the O-APs and PDs are vertically oriented, we have $\psi_m=\varphi_m$.

Suppose the effective area of a PD is $A_\text{unit}$, the received light intensity of the $k$-th PD from the $m$-th O-AP is
\begin{eqnarray}\label{e:I^rx}
&&\hspace{-1cm} I_{m,k}^{\text{rx}} \left(x_{P,k},y_{P,k},\varepsilon, \xi_m, R(\varphi_m)\right) \nonumber\\
&&\hspace{-1.4cm} = I^{\text{tx}}\!\cdot\! \underbrace{ \int{\frac{ m_0 + 1 }{ 2\pi }\! \cos^{m_0} \!\!\big( \phi_{ m }\left( \varphi_m, \lambda \right) \!\big) \chi(\lambda) d\lambda } }_{R\left( \varphi_m \right)} \!\cdot\!\underbrace{\frac{A_{\text{unit}} \cos{\psi_m}}{d_{m,k}^2}}_{\Omega\left( \varepsilon,\xi_m \right)},
\end{eqnarray}
where 
\begin{itemize}[leftmargin=0.55cm]
    \item $I^{\text{tx}}$ is the total light intensity emitted from the LED. Without loss of generality, we set $I^{\text{tx}}=1$.
    \item $R\left( \varphi_m \right)$ is the radiation pattern of the LED considering all wavelengths emitted by a single light source, and $\chi(\lambda)$ is the distribution of wavelength $\lambda$ with $\int{ \chi(\lambda) d\lambda } = 1$\footnote{Practical optical sources cannot produce light of a single frequency, as illustrated in Fig.~\ref{f:Obf}(b).}.
    $m_0\allowbreak= \allowbreak-\frac{1}{\log_2\left(\cos\Phi_{1/2}\right)}$ is Lambert's mode number, and we set $m_0=1$ as the typical half-power angle is $\pi/3$.
    In the absence of the collimating lens, we have $\phi_m \big( \varphi_m, \lambda \big) = \varphi_m$, and $R\left( \varphi_m \right)$ can be refined as $R\left( \varphi_m \right) = \frac{ 1 }{\pi } \cos \varphi_m$. 
    \item $\Omega \left( \psi_{m} \right)$ is the solid angle, in which $A_\text{unit}$ is the unit area both for optical communication and optical sensing, and $d_{m,k}$ represents the distance between the $m$-th O-AP and the $k$-th PD $(x_{P,k},\allowbreak y_{P,k},\allowbreak z_{P,k})$.
\end{itemize}

\subsection{Optical communication}

Unlike radio communication, phase modulation and coherent detection in optical communication are very expensive to realize, as it is challenging to match the frequency and polarization of the local laser with that of the incoming optical signal, or even impossible due to the incoherent light emitted by light sources such as LED \cite{wang2017visible}.
The modulation and demodulation schemes that find wide applications in optical communication systems are IM/DD.

Consider an Orthogonal Frequency Division Multiplexing (OFDM)-enabled O-ISAC system with $N$ orthogonal subcarriers.
Let $\bm{U}=\left[\bm{u}_1 , \bm{u}_2 , ... , \bm{u}_L \right] \in \mathbb{C}^{(N/2-1)\times L}$ 
be a matrix to be sent to the devices.
For intensity modulation, we adopt the DC-biased optical OFDM (DCO-OFDM) strategy, which involves two key steps: 1) applying Hermitian symmetry \cite{afgani2006visible} to the inputs of the Inverse Discrete Fourier Transform (IDFT). That is, we construct $\bm{X}=\left[\bm{x}_1 , \bm{x}_2 , ... , \bm{x}_L \right] \in \mathbb{C}^{N\times L}$, where
\begin{eqnarray}\label{e:x}
    &&\hspace{-1.2cm} \bm{x}_\ell\!=\![0,u_{1,\ell},u_{2,\ell},...,u_{N/2-1,\ell},0,u_{N/2-1,\ell}^*,...,u_{2,\ell}^*,u_{1,\ell}^*]^\top\!\!,
\end{eqnarray}
$\ell =1, 2, ..., L$, and perform an $N$-point IDFT yielding real-valued OFDM samples $\bm{V}\in \mathbb{R}^{N\times L}$. 2) Introducing a DC bias in the LED's driving circuit to ensure the signal remains non-negative.
Recall that we use lowercase bold letters to represent the column-wise vectorized form of a matrix, e.g., 
$\bm{x} \triangleq \emph{vec} \left( \bm{X} \right)$ 
and $\bm{v} \triangleq \emph{vec} \left( \bm{V} \right)$.
The transformation from $\bm{x}$ to $\bm{v}$ is 
\begin{equation}\label{e:v}
\bm{v}=\frac{1}{\sqrt N} \left(\bm{I}_L \otimes \bm{F}_N^H\right)\bm{x},
\end{equation}
where 
$\bm{F}_N$ denotes the $N$-point discrete
Fourier transform (DFT) matrix ($\bm{F}_N^H$ is the IDFT matrix), 
$\otimes$ is the Kronecker product, and
$\bm{I}_L$ is the $L$-dimensional identity matrix.
The transmitted signal after pulse shaping can be written as \cite{FL}
\begin{equation}
s (t) = \sum_{\ell = 0} ^ {L-1} \sum_{n = 0} ^ {N-1} {\bm{v}[n + \ell N] g(t - n T_\text{sam} - \ell T_\text{sym})} ,
\end{equation}
where $g(\cdot)$ is the shaping pulse,
$T_\text{sam}$ is the sample duration, and
$T_\text{sym}$ is the OFDM symbol duration, $T_\text{sym} = N T_\text{sam}$.

Without loss of generality, we consider one device in the system.
The signal received by the $k$-th ($k = 1,2,...,\kappa$) PD can be written as
\begin{equation}\label{e:r}
\begin{split}
 r_{k}(t) 
 &= \sum_{m=1} ^{\mu} {s(t) \ast \left[ h_{m,k} \delta\left( t - \frac{d_{m,k}}{c}\right)\right] } + w_k (t) \\
 &= \sum_{m=1} ^{\mu} {s\left(t - \frac{d_{m,k}}{c}\right) h_{m,k}} + w_k (t),
\end{split}
\end{equation}
where 
$d_{m,k}$ denotes the distance between the $m$-th O-AP and the $k$-th device, and
$w_k (t) $ is additive white Gaussian noise (AWGN) added to the electrical domain signal \cite{ghassemlooy2019optical}.
According to \eqref{e:I^rx},
the channel gain between the $m$-th O-AP and the $k$-th PD can be written as
\begin{eqnarray}\label{e:h_nt}
&&\hspace{-0.7cm} h_{m,k}
=\!\frac{I_{m,k}^{\text{rx}}}{I^{\text{tx}}}
\overset{(a)}{=} \frac{ A_{\text{unit}}}{\pi d_{m,k}^2} \cos \varphi_{m} \cos \psi_{m} 
= \frac{ A_{\text{unit}} \mathcal{H}^2 }{\pi d_{m,k}^4} ,
\end{eqnarray}
where (a) follows from \eqref{e:I^rx} without optical beamforming.
At the receiver, we sample $r_k(t)$ and perform DFT, yielding
\begin{align}\label{e:DFT_y}
\bm{y}_k [n+\ell N] 
= \bm{x}[n + \ell N] \bm{\Delta}_k^\top \bm{h}_k +\bm{w}_k [n+\ell N],
\end{align}
where 
$\bm{\Delta}_k = \big[ e^{-j2 \pi \frac{n}{N} \frac{d_{1,k}}{cT_\text{sam}}} , 
e^{-j2 \pi \frac{n}{N} \frac{d_{2,k}}{cT_\text{sam}}} ,\allowbreak ...,\allowbreak e^{-j2 \pi \frac{n}{N} \frac{d_{\mu,k}}{cT_\text{sam}}} \big]^\top$ is a phase matrix,
$\bm{h}_k = \left[ h_{1,k},\allowbreak h_{2,k},\allowbreak ..., \allowbreak h_{\mu,k}\right]^\top$ is the channel coefficient vector, and
$\bm{w}_k [n+\ell N] \sim \mathcal{N} (0, \sigma ^2)$.
Compared to radio frequency communication, IM/DD exclusively relies on measuring the intensity of light rather than its phase. Consequently, optical carriers do not introduce extra phase shifts at the baseband due to time offsets.

Overall, the signal received by all $\kappa$ PDs can be written in a compact form as
\begin{eqnarray}\label{e:PD-array_h}
\bm{y} \!=\!
\bm{x} \begin{bmatrix} \bm{\Delta}_1^\top \bm{h}_1 & \bm{\Delta}_2^\top \bm{h}_2 & ... & \bm{\Delta}_{\kappa}^\top \bm{h}_{\kappa} \end{bmatrix} + \begin{bmatrix} \bm{w}_1 & ... & \bm{w}_{\kappa} \end{bmatrix}\!. 
\end{eqnarray}
To decode the transmitted signal, we employ the maximum ratio combining (MRC) \cite{PNC}:
\begin{eqnarray}\label{e:PD-array_h_prime}
    \bm{y}_{\text{MRC}} 
\hspace{-0.2cm}&=\hspace{-0.2cm}& 
\bm{x} \sum_{k=1}^\kappa{\| \bm{\Delta}_{k}^\top \bm{h}_{k}\|^2} + \sum_{k=1}^\kappa{\| \bm{w}_{k}\bm{\Delta}_{k}^\top \bm{h}_{k}\|^2},
\end{eqnarray}
from which an estimate of $\bm{x}$ is given by
\begin{align}\label{e:estimation_x}
\widehat{\bm{x}} = 
\frac{\bm{y}_{\text{MRC}} }{\sum_{k=1}^\kappa{\| \bm{\Delta}_{k}^\top \bm{h}_{k}\|^2}}.
\end{align}
With $\widehat{\bm{X}} \triangleq \emph{devec} (\widehat{\bm{x}})$, the transmitted message can be reconstructed as
\begin{align}
\widehat{\bm{U}}[n,\ell] =\frac{1}{2} \left( \widehat{\bm{X}}[n+1,\ell] + \widehat{\bm{X}}^*[N+1-n,\ell] \right).
\end{align}

\begin{table*}[!t]
\centering
\caption{A summary of the coordinate systems.}
\setlength{\tabcolsep}{5mm}
\begin{tabular}{c|c|c}
\hline
\hline
\textbf{Coordinate Systems} & \textbf{Objects} &\textbf{Coordinates}\\
\hline
\multirow{3}{*}{ The 3D real-world coordinate system in Fig.~\ref{f:model} } 
& The $m$-th LED and pinhole
& $\bm{p}_{\text{OAP},m}=(\varepsilon \cos{\xi_m}, \varepsilon \sin{\xi_m}, \mathcal{H})$\\ \cline{2-3}
& The $k$-th PD of the target device & $\bm{p}_{P,k}=(x_{P,k},y_{P,k},z_{P,k} )$\\ \cline{2-3}
& The target device & $\bm{p}_D = (x_D,y_D, z_D )$\\ \hline
The 3D camera coordinate system of the $m$-th O-AP
& The target device
& $\bm{p}_{C,m}=(x_{C,m},y_{C,m},z_{C,m} )$\\ \hline
The 2D film plane coordinate system of the $m$-th O-AP
& The target device
& $\bm{p}_m=(x_m,y_m )$\\ \hline \hline
\end{tabular}
\label{t-coordinate}
\end{table*}

\subsection{Optical sensing\label{subs:O_sensing}}

Optical sensing aims to detect and measure various physical, chemical, or biological parameters, such as position, temperature, pressure, and medical diagnostics.
In this paper, we focus on positional sensing to estimate the three-dimensional (3D) coordinate of a target device.

In RF-ISAC systems \cite{liu2020joint}, a critical assumption is that there exists only a finite number of scatters in the environment, hence the number of echoes is small. This is because excessive echo interference can overpower the desired signal, resulting in a low signal-to-interference-and-noise ratio.
In contrast, this paper considers a more realistic setup, where all objects in the environment can be reflectors and the reflected light from all range bins can be collected by the optical sensor.

\begin{figure}
 \centering
 \includegraphics[width=0.8\linewidth]{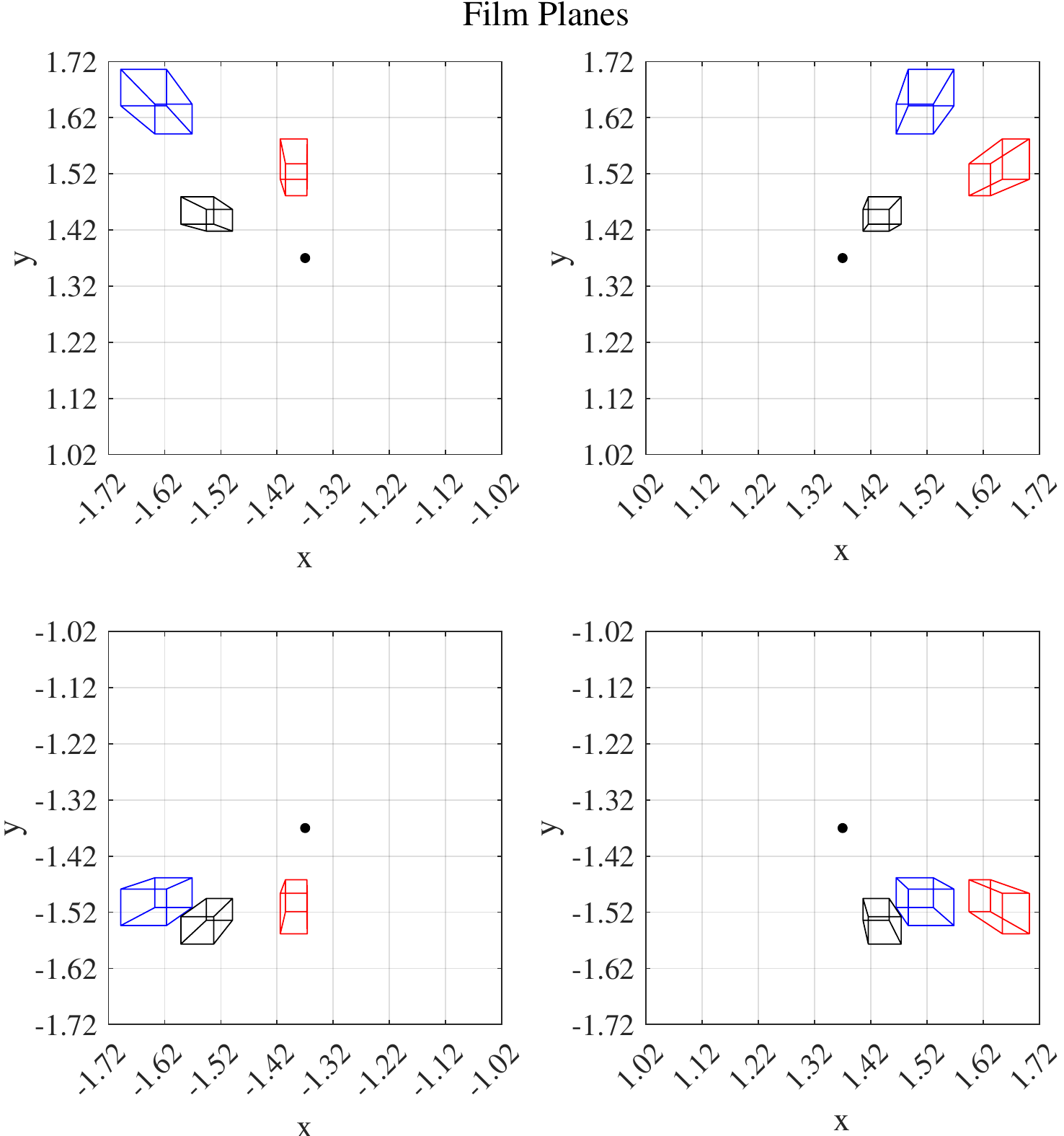}
 \caption{An illustration of the film planes captured by the pinhole cameras of four optical O-APs. Each file plane has a 2D coordinate system.}
 \label{f:imaging}
\end{figure}

Unlike RF-ISAC which analyzes the composition of the received signal to extract sensing information, we utilize the pinhole imaging principle to map all reflected light onto a film plane for sensing.
Thanks to the ultra-high frequency characteristics of the optical signal, the undesired reflected light is separated from the desired reflections as they are mapped to distinct pixel locations on the film plane, effectively avoiding interference.
An illustration is given in Fig.~\ref{f:imaging}, in which the pinhole cameras of the O-APs capture $\mu$ images of the environment from $\mu$ perspectives.

Before exploring the optical sensing operations, let us clarify the relative coordinate systems in the system model.
As summarized in Table \ref{t-coordinate}, the coordinates of both LEDs and PDs defined earlier are based on the real-world coordinate system. 
We assume the PD array is placed at the center of the target device. Therefore, in the real-world coordinate system, the coordinate of the target device, denoted by $\bm{p}_D = (x_D, y_D, z_D)$, corresponds to the center of the PD array.

The pinhole camera of each O-AP also maintains a 3D coordinate system, called the camera coordinate system, wherein the pinhole is the origin with the coordinate $(0,0,0)$. Without loss of generality, we design the $\mu$ camera coordinate systems such that they share the same orientation. In other words, the 3D coordinate of the target device in the $m$-th camera coordinate system $\bm{p}_{C,m} = (x_{C,m}, y_{C,m}, z_{C,m})$ satisfies
\begin{eqnarray}\label{e:relative}
&&\hspace{-1.3cm} x_{C,m} \!=\!x_{C,1} + \varrho_{x,m},~   y_{C,m}\!=\!y_{C,1} + \varrho_{y,m}, ~  z_{C,m}\!=\! z_{C,1},
\end{eqnarray}
where $\varrho_{x,m}$ and $\varrho_{y,m}$ are constants since the relative positions among cameras are fixed. The above transformations facilitate us to represent $\bm{p}_{C,m}$, $\forall m$ using $\bm{p}_{C,1}$.
For the target device, the coordinates $\bm{p}_{C,m}$ (under the camera coordinate system of the $m$-th O-AP) and $\bm{p}_D$ (under the real-world coordinate system) can be transformed to each other via \cite{do2016depth}
\begin{equation}\label{e:W_C_transform}
\begin{bmatrix} \bm{p}_{C,m}\\ 1 \end{bmatrix} 
= \begin{bmatrix} \bm{Q}_m & \bm{t}_m\\ 0 & 1 \end{bmatrix} \begin{bmatrix} \bm{p}_D\\ 1 \end{bmatrix},
\end{equation}
where 
$\{\bm{Q}_m:m=1,2,...,\mu\}$ are $3\times3$ rotation matrices,
$\{\bm{t}_m:m=1,2,...,\mu\}$ are $3\times1$ positional vectors,
and $\{\bm{Q}_m \}$ and $\{\bm{t}_m \}$ denote the exterior orientation parameters (EOPs).

Each film plane has a 2D plane coordinate system.
We denote by $\bm{p}_m = (x_m, y_m)$ the coordinate of the target device on the 2D plane coordinate systems. According to the pinhole imaging principle, $\bm{p}_m$ can be obtained from either $\bm{p}_{C,m}$ or $\bm{p}_{D}$.
Their relationships can be written as 
\begin{align} \label{e:position}
z_{C,m} \begin{bmatrix} \bm{p}_m\\ 1 \end{bmatrix} 
= \begin{bmatrix} f_{x,m} & 0 & 0 \\ 0 & f_{y,m} & 0 \\ 0 & 0 & 1 \end{bmatrix}  \bm{p}_{C,m} 
 \triangleq \bm{K}_m \bm{p}_{C,m}, 
\end{align}
where $f_{x,m}$ and $f_{y,m}$ are interior parameters (focal lengths) of the pinhole camera and $\bm{K}_m$ denotes the interior orientation parameters (IOPs). 
In particular, we assume the IOPs of the cameras are the same and define $\bm{K}_m \triangleq \bm{K}$.

Given the sensed images, we perform image processing algorithms to estimate the coordinates of the target.
The estimation accuracy depends on both the light intensity and the contrast ratio. Since the received light intensity is contaminated by AWGN, the coordinate estimation error follows Gaussian distributions when the pixel size is sufficiently small. Therefore, we can write the estimated coordinates as
\begin{align}\label{e:position_hat}
\widehat{\bm{p}}_m 
=\bm{p}_m + \bm{e}_{m} 
=\begin{bmatrix} x_m\\ y_m \end{bmatrix} + \begin{bmatrix} e_{x,m}\\ e_{y,m} \end{bmatrix},
\end{align}
where the variance of the estimation error is inversely propositional to the light intensity, i.e., $e_{x,m}, e_{y,m} \sim\mathcal {N} (0,\eta \frac{\sigma_I^2}{I_m^{\text{ref}}})$\footnote{The detection of pixel intensity on the film plane follows a Gaussian distribution \cite{jain1989fundamentals}. As a result, the coordinates derived from this process also adhere to a Gaussian distribution centered around their actual positions. This outcome stems from the mapping of intensity to coordinates, approximated through a first-order Taylor expansion, effectively rendering it a linear combination of Gaussian variables.}, and
$\eta$ is a scaling factor determined by the related size of the film plane to the environment and the distance of the film plane to the pinhole; 
$\sigma_I^2$ is the variance of AWGN in the received light;
$I_m^{\text{ref}}$ is the reflected light intensity given by
\begin{align}\label{eq:sensed}
I_{m,k}^{\text{ref}} = \! \left[\sum_{m=1}^{\mu} {I_{m,k}^{\text{rx}} }\right] \!\cdot
\rho_{\text{ref}} A_{\text{unit}} \frac{ \cos\varphi_{m}}{d_{m,k}^2},
\end{align}
where $I_{m,k}^{\text{rx}}$ is defined in \eqref{e:I^rx}, and the first term (i.e., the summation) represents the superposition of the intensities of all light sources on the reflectors;
$\rho_{\text{ref}}$ is the reflection coefficient of the reflector;
$A_{\text{unit}}$ represents the area of the reflector;
$\varphi_{m}$ is the angle of incidence (AoI) of the reflected signal, which is equal to the AoD of the $m$-th O-AP.
Eq. \eqref{e:I^rx} and \eqref{eq:sensed} also highlight the synergy between optical sensing and optical communication. Our proposed optical ISAC system is designed to utilize data transmission signals for sensing purposes, addressing the absence of inherent sensing signals. This innovative approach negates the necessity for additional light sources and circumvents reliance on ambient light, streamlining the system for enhanced efficiency.

From \eqref{e:relative}, \eqref{e:position} and \eqref{e:position_hat}, we have
\begin{align}\label{eq:solve_z}
z_{C,1}\!\! \begin{bmatrix} \widehat{\bm{p}}_m \\ 1 \end{bmatrix} 
\!=\! \bm{K} \bm{p}_{C,m} \!+\! \begin{bmatrix} \bm{e}_m \\ 1 \end{bmatrix}
\!=\! \bm{K}& \left( \bm{p}_{C,1} \!+\! \bm{v}_{m} \right) \!+\! \begin{bmatrix} \bm{e}_m \\ 1 \end{bmatrix}, 
\end{align}
where $m=1,2,...,\mu $.
After some manipulations, \eqref{eq:solve_z} can be reorganized into a more compact form as
\begin{equation}\label{e:MSE_fuc}
\begin{split}
 &\begin{bmatrix} 0 \\ 0 \\ ...\\-f_x \varrho_{x,m}\\-f_y \varrho_{y,m}\\ ... \end{bmatrix}
 \!=\! \begin{bmatrix} 
 f_x & 0 & -\widehat{x}_1 + e_{x,1} \\
 0 & f_y & -\widehat{y}_1 + e_{y,1} \\
 ... & ... & ... \\
 f_x & 0 & -\widehat{x}_m + e_{x,m} \\
 0 & f_y & -\widehat{y}_m + e_{y,m} \\
 ... & ... & ... \end{bmatrix}
 \begin{bmatrix} x_{C,1} \\ y_{C,1} \\ z_{C,1} \end{bmatrix}
 \\
 &\triangleq\hspace{0.5cm} \bm{\gamma}=\bm{\Sigma} \bm{p}_{C,1}.
\end{split}
\end{equation} 
Note that $\bm{\gamma}$ is a constant vector. From the sensed $\widehat{\bm{p}}_m$ in \eqref{e:position_hat}, we can construct an estimated $\widehat{\bm{\Sigma}}$ as
$$\widehat{\bm{\Sigma}} = 
\begin{bmatrix} 
 f_x & 0 & -\widehat{x}_1 \\
 0 & f_y & -\widehat{y}_1 \\
 ... & ... & ... \\
 f_x & 0 & -\widehat{x}_m \\
 0 & f_y & -\widehat{y}_m \\
 ... & ... & ... \end{bmatrix}.$$
Then, $\widehat{\bm{p}}_{C,1}$ is estimated by
\begin{equation} \label{e:P_C_hat}
 \widehat{\bm{p}}_{C,1} = 
 (\widehat{\bm{\Sigma}}^\top \widehat{\bm{\Sigma}})^{-1} \widehat{\bm{\Sigma}} ^\top \bm{\gamma}.
\end{equation}
Finally, the 3D coordinates of the target device in the real-world coordinate system are calculated from \eqref{e:W_C_transform} as
\begin{equation}
\begin{bmatrix} \widehat{\bm{p}}_D\\ 1 \end{bmatrix} ={\begin{bmatrix} \bm{Q}_1 & \bm{t}_1\\ 0 & 1 \end{bmatrix}}^{-1}\begin{bmatrix} \widehat{\bm{p}}_{C,1} \\ 1 \end{bmatrix}.
\end{equation}
We use the MSE of the coordinates to measure the sensing accuracy, giving
\begin{equation}\label{e:MSE}
\text{MSE}_P = \mathbb{E}\left\{{\Vert \widehat{\bm{p}}_D - \bm{p}_D \Vert}^2\right\}.
\end{equation}

It is worth noting that 1) there must be at least 2 pinhole cameras to ensure that the row rank of matrix $\bm{\Sigma}$ in \eqref{e:MSE_fuc} is larger than the 3; 2) A simple expansion of the matrices in \eqref{e:MSE_fuc} and \eqref{e:P_C_hat} enables simultaneous estimation of multiple targets at arbitrary positions, including those at different heights, as illustrated in Fig.~\ref{f:imaging}. This matches our intuition, as each camera observes two-dimensional information, necessitating at least two or more cameras to provide sufficient degrees of freedom for estimating positions in the 3D space.

\section{Synergies between Optical Sensing and Optical Communication}\label{sec:III}
Sec. \ref{sec:II} details the signal processing of both sensing and communication in our O-ISAC system.
Expanding upon the signal flow, this section will uncover the connections between the factors that impact the performance of optical sensing and optical communication.
Through the exploration, we will unite these two processes seamlessly, forming a two-phase O-ISAC operation mechanism. 
Then, we shall turn our attention to optimizing the arrangement of light sources within the room. This strategic optimization promises to boost the effectiveness of the first phase of O-ISAC.

\subsection{Synergies between optical sensing and optical communication and a two-phase operation mechanism} 

Drawing from \eqref{e:PD-array_h_prime}, we can express the SNR in optical communication as
\begin{eqnarray}
    \text{SNR} = \frac{\sum_{k=1}^\kappa{\| \bm{\Delta}_{k}^\top \bm{h}_{k}\|^2}}{\sigma^2}.
\end{eqnarray}
Thus, the decoding performance of optical communication is reliant on the equivalent channel gain at the baseband $\sum_{k=1}^\kappa{\| \bm{\Delta}_{k}^\top \bm{h}_{k}\|^2}$.
In particular, it can be further refined as
\begin{eqnarray} \label{e:equivalent_channel}
&&\hspace{-0.2cm}\sum_{k=1}^\kappa{\| \bm{\Delta}_{k}^\top \bm{h}_{k}\|^2}\nonumber\\
&&\hspace{-0.6cm} = \sum_{k=1}^\kappa{\left[ \sum_{m=1}^\mu e^{-j 2 \pi \frac{n}{N} \frac{d_{m,k}}{c T_\text{sam}}} h_{m,k} \right] \left[ \sum_{m=1}^\mu e^{-j 2 \pi \frac{n}{N} \frac{d_{m,k}}{c T_\text{sam}}} h_{m,k} \right]^* }\nonumber\\
&&\hspace{-0.7cm} \overset{(a)}{\approx} \sum_{k=1}^\kappa{\left[ e^{-j 2 \pi \frac{n}{N} \frac{d_{1,k}}{c T_\text{sam}}} \sum_{m=1}^\mu h_{m,k} \right] \left[ e^{-j 2 \pi \frac{n}{N} \frac{d_{1,k}}{c T_\text{sam}}} \sum_{m=1}^\mu h_{m,k} \right]^* }\nonumber\\
&&\hspace{-0.6cm} = \sum_{k=1}^\kappa{\left[  \sum_{m=1}^\mu h_{m,k} \right]^2 }
\overset{(b)}{=} \sum_{k=1}^\kappa{\left[  \sum_{m=1}^\mu I_{m,k}^{\text{rx}} \right]^2 } .
\end{eqnarray}
where (a) follows thanks to IM/DD -- the receiver measures the amplitude, as opposed to the phase, of the received signal for decoding.
As a result, only the baseband subcarrier can cause phase shifts on the received symbols.
Since the time delays are relatively small compared with the OFDM symbol duration, the additional phase shifts on different subcarriers are approximately the same.
(b) follows from \eqref{e:h_nt}.
Overall, the equivalent channel gain in optical communication is proportional to the received light intensity at each PD $\sum_{m=1}^\mu \! {I_{m,k}^{\text{rx}}}$.

When it comes to optical sensing, on the other hand, the performance of optical sensing relies also on the intensity of the light received by the target PD, as indicated in the first term of \eqref{e:position_hat}. 
This underscores the interconnected nature of optical communication and optical sensing. Essentially, we can work towards optimizing the O-ISAC system with a shared objective, i.e., $\sum_{m=1}^\mu \! {I_{m,k}^{\text{rx}}}$, without sacrificing the performance of either the optical communication or optical sensing aspects. This synergy between our goals makes the optimization process more efficient and beneficial for the overall functionality of O-ISAC.

\begin{figure}
    \centering
    \includegraphics[width=1\linewidth]{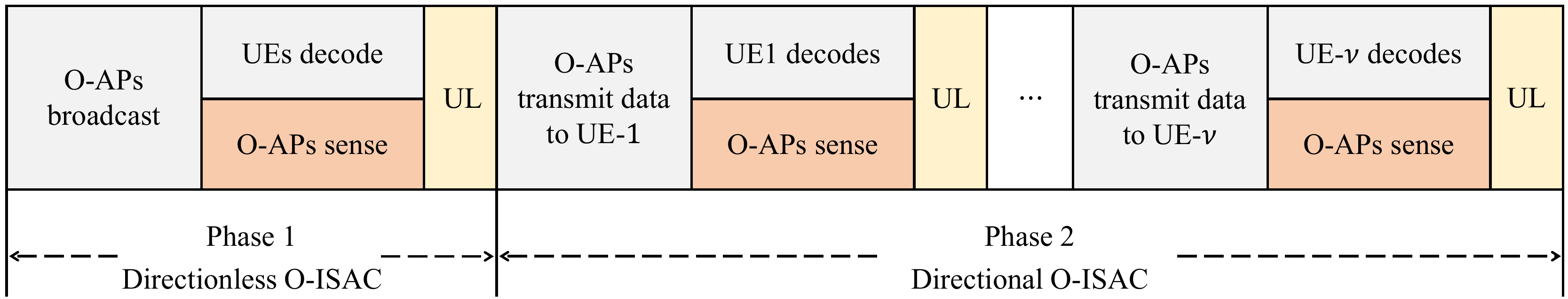}
        \caption{The workflow of the proposed two-phase O-ISAC system.}
    \label{f:workflow}
\end{figure}

In $I_{m,k}^{\text{rx}}\allowbreak\left(x_{P,k},\allowbreak y_{P,k},\allowbreak \varepsilon,\allowbreak \xi_m,\allowbreak R(\varphi_m)\right)$,
$x_{P,k}$ and $y_{P,k}$ are the devices' position, which is uncontrollable, 
$\varepsilon$ and $\xi_m$ are the O-AP coordinates,
and $R(\varphi_m)$ is the radiation pattern of the light source.
Therefore, we can maximize $\sum_{m=1}^\mu \! {I_{m,k}^{\text{rx}}}$ by finding the optimal $\varepsilon$, $\xi_m$, and $R(\varphi_m)$. Specifically, optimizing the light source distribution is crucial for system performance in broadcast scenarios, enabling the detection of a broader range of devices; on the other hand, concentrating light intensity is suitable for scenarios targeting specific devices. In this context, we can divide the O-ISAC system into two operation phases, the workflows of which are illustrated in Fig.~\ref{f:workflow}.

\begin{itemize}
    \item \textbf{Phase 1} (directionless O-ISAC): The O-APs broadcast a control message to all devices periodically in Phase 1, and sense the devices' states globally based on the reflected light.
    In this phase, the performance of the system within the region is primarily determined by the light source layout. In Section \ref{subs:layout}, we will present specific optimization strategies to enhance the area within the room where the superimposed optical intensity exceeds a predefined threshold.
    \item \textbf{Phase 2} (directional O-ISAC): The O-APs serve the devices in a TDMA fashion. 
    Given the sensed states in the first phase, the communications between the O-APs and devices are improved by {\it optical beamforming}, a scheme to adjust the emitted light intensity distribution of the optical source. In phase 2, our optimization objective is to improve the radiation pattern $R(\varphi_m)$ of the light sources in order to achieve the convergence of optical intensity across all O-APs and concentrate it on the target object. The O-APs further utilize the enhanced reflected light of phase 2 to accurately sense and track the devices' states. The details will be presented later in Section \ref{sec:IV}.
\end{itemize}

\subsection{Source layout optimization\label{subs:layout}}

In the first phase of O-ISAC, the received light intensity is determined by the distribution of optical sources. 
In traditional optical communication endeavors, a commonly employed optimization criterion for light source distribution is uniformity\cite{deqiang2007optimal,zuo2022symmetrical,liu2017optimization}, where the objective is to achieve the most even light intensity across a given plane.
More rigorously, such metrics can be quantified as the MSE of the received signal strength.
The goal is to find the optimal LED distribution to minimize the MSE of the signal strength:
\begin{eqnarray}\label{e:Intensity_MSE}
    &&\hspace{-1.3cm} \min_{\varepsilon, \{\xi_m \}}  \frac{1}{\mathcal{W}\mathcal{L}}\int_{- \frac{\mathcal{L}}{2}}^{ \frac{\mathcal{L}}{2}}{\!\int_{- \frac{\mathcal{W}}{2}}^{ \frac{\mathcal{W}}{2}}{\! \left[{\sum\limits_{m} {I_{m,k} ^ { \text{rx} } } } \!-\! \mathbb{E}\left\{{ \sum\limits_{m} {I_{m,k}^{\text{rx}}}}\right\}\right]^2 \!dx }dy }.
\end{eqnarray}
Optimizing \eqref{e:Intensity_MSE} enables the device to receive a consistently stable optical signal while in motion within the room.

In this paper, we put forth a distinct optimization criterion. Firstly, we establish a threshold $\rho_I$ that corresponds to an acceptable communication and sensing performance. Our optimization objective is to maximize the extent of the area within the room where the received signal strength surpasses the threshold. Due to the path loss of light, comparisons can only be made for distributions on the same plane. Since the ground level experiences the most significant light intensity attenuation within a room, optimizations are conducted at ground level for simplicity and without loss of generality, i.e., $z_{P,k} = 0$. By substituting $\mathcal{H}-z_{P,k}$ for $\mathcal{H}$, expressions for various altitudes can be derived. Overall, our proposed system is designed to serve devices at any altitude, accommodating those positioned at differing heights simultaneously. Specifically, we formulate (P1):
\begin{subequations}\label{e:optimization_h}\begin{align}
\max_{\varepsilon, \{\xi_m \}}& \frac{1}{\mathcal{W}\mathcal{L}}\! \int_{- \frac{\mathcal{L}}{2}}^{ \frac{\mathcal{L}}{2}}\!{\int_{- \frac{\mathcal{W}}{2}}^{ \frac{\mathcal{W}}{2}}\!{ \frac{1}{2} \Big[ \text{sgn}\big( \sum\limits_{m} {I_{m,k} ^ { \text{rx} } } \!-\! \rho_I \big)\! +\!1 \Big]dx }dy },\label{e:P1}
\\ {\text {s.t.}~} 
    &I_{m,k}^{\text{rx}} = R(\varphi_m)\cdot\frac{ A_{\text{unit}} \cos{\psi_m} }{  d_{m,k}^2} ,\label{e:P1I}
    \\&  R(\varphi_m) = \frac{1}{\pi} \cos{\varphi_m},\label{e:P1R}
  \\ d_{m,k} \!&=\!\! \sqrt{\!(\varepsilon \cos{ \xi_m} \!-\! x_{P,k})^2 \!+\! (\varepsilon \sin{ \xi_m} \!-\! y_{P,k})^2 \!+\! \mathcal{H}^2}, \label{e:P1d}
  \\&  0\leq\varepsilon\leq \min{\{\mathcal{W},\mathcal{L}\}},\label{e:P1ve}
  \\& 0\leq\xi_m < 2\pi,~m=1,2,...,\mu.\label{e:P1xi}
     \end{align}  \end{subequations} 
As can be seen, rather than uniformly compromising communication across all regions, our metric lies in enabling acceptable light intensity from a broader range of regions. 

\begin{thm}\label{thm:dist}
In the first phase of O-ISAC, the optimal source layout that maximizes the proportion of areas exceeding the threshold $\rho_I$, i.e., the optimal solution to (P1), can be approximated by $\bm{p}_{\text{O-AP},m}^* = \big(\varepsilon^* \cos{\xi_m^*},\allowbreak \varepsilon^* \sin{\xi_m^*},\allowbreak \mathcal{H} \big)$, where
 \begin{eqnarray}
 &&\hspace{-1cm}\label{e:varepsilon*} \varepsilon^* = \sqrt{\frac{\sqrt{ \frac{5 A_\text{unit} \mathcal{H}^2 }{2\pi \rho_I}} - \mathcal{H}^2}{\tan^2{\frac{\pi}{\rho_I}}}} , \quad
 \xi_m^* = \frac{2\pi (m-1)}{\mu} +  \frac{\pi}{4}.
 \end{eqnarray}
\end{thm}

We next prove Theorem \ref{thm:dist}.
To find the optimal value $\varepsilon^*$, we first discuss the critical condition in \eqref{e:P1}. Let
\begin{eqnarray}\label{e:proof_varepsilon}
&&\hspace{-1cm}\rho_I = \sum\limits_{m} {I_{m,k} ^ { \text{rx} } } 
\overset{(a)}{=} \sum_{m=1}^\mu \frac{ A_{\text{unit}}}{\pi} \frac{ \mathcal{H}^2 }{d_{m,k}^4} \overset{(b)}{=} \sum_{m=1}^\mu\Biggl[ \frac{ A_{\text{unit}}}{\pi} \cdot\\
&&\hspace{1cm}\frac{ \mathcal{H}^2 }{\left( (\varepsilon \cos{\xi_m}\! -\! x)^2 \!+\! (\varepsilon \sin{\xi_m} \!-\! y)^2 \!+\! \mathcal{H}^2 \right)^2}\Biggr] , \nonumber
\end{eqnarray}
where 
(a) follows from \eqref{e:h_nt},
and (b) follows from \eqref{e:P1d}.
Analytically solving \eqref{e:proof_varepsilon} is quite challenging.
The difficulty arises from the intricate complexities in determining the light intensity distribution when multiple distributed light signals intersect at a target point (i.e., the summation over $m$).
To meet the analytical challenges, we make two assumptions: 1) all LEDs are uniformly distributed on the circle;
2) the area that exceeds the threshold is maximized when the critical point on the symmetry axis between adjacent light sources is farthest from the origin.
The accuracy of the assumptions will be confirmed later through simulations.

From the first assumption, we can assume an initial phase angle of $\xi_m = \frac{\pi}{\mu}(2m-1)$, meaning that the first and $\mu$-th O-APs are symmetrically distributed about the x-axis, with the positive x-axis serving as the symmetry axis for the first and $\mu$-th O-APs.
Since the intensity of light rapidly attenuates with distance, we can only consider the impact of adjacent LEDs. Assuming that the two adjacent LEDs provide the majority of the light intensity at the critical point on the positive x-axis, we can derive the following relationship:
\begin{eqnarray}\label{e:proof_varepsilon1}
&&\hspace{-1.2cm} \frac{4}{5}\rho_I \!\approx \!\!\!\!\sum\limits_{m=1,\mu}\!\!\!\! {I_{m,k} ^ { \text{rx} } } =\!\!\!\! \sum_{m=1,\mu}\!\! \!\!\frac{ A_{\text{unit}}}{\pi} \frac{ \mathcal{H}^2 }{(\varepsilon^2 + \mathcal{H}^2 + x^2 - 2\varepsilon\cos{\frac{\pi}{\mu}x})^2} .
\end{eqnarray}

The first and $\mu$-th O-APs are symmetric. Considering the light intensity brought by one of them, we get
\begin{eqnarray}\label{e:proof_varepsilon2}
&&\hspace{-0.75cm} \frac{2}{5} \rho_I \approx  \frac{ A_{\text{unit}}}{\pi} \frac{ \mathcal{H}^2 }{(\varepsilon^2 + \mathcal{H}^2 + x^2 - 2\varepsilon\cos{\frac{\pi}{\mu}x})^2} .
\end{eqnarray}
Next, we need to find the value of $\varepsilon$ that maximizes $x$. Taking the partial derivative of both sides with respect to (w.r.t.) $\varepsilon$, we obtain
\begin{eqnarray}\label{e:proof_varepsilon3}
&&\hspace{-0.75cm} 0=  \frac{-2\Big( 2\varepsilon + 2x \frac{\partial}{\partial\varepsilon}x - 2 \cos{\frac{\pi}{\mu}}x -2\varepsilon\cos{\frac{\pi}{\mu}}\frac{\partial}{\partial\varepsilon}x \Big) }{(\varepsilon^2 + \mathcal{H}^2 + x^2 - 2\varepsilon\cos{\frac{\pi}{\mu}x})^3} .
\end{eqnarray}
Substituting $\frac{\partial}{\partial\varepsilon}x = 0$ into \eqref{e:proof_varepsilon3}, we get:
\begin{eqnarray}\label{e:proof_varepsilon4}
&&\hspace{-0.75cm} 0=  \frac{\varepsilon - \cos{\frac{\pi}{\mu} x} }{(\varepsilon^2 + \mathcal{H}^2 + x^2 - 2\varepsilon\cos{\frac{\pi}{\mu}x})^3} .
\end{eqnarray}

Combining \eqref{e:proof_varepsilon4} and \eqref{e:proof_varepsilon2}, the optical $\varepsilon^*$ can be solved, yielding \eqref{e:varepsilon*}.

Then, we proceed to solve for the optimal phase angle $\xi_m^*$. 
Under the assumption that all LEDs are uniformly distributed on the circle, we can define an initial phase angle as $\xi_0 = \xi_m - \frac{2\pi m}{\mu}$. Thus, the next step is to determine the optimal initial phase angle $\xi_0^*$.
Similar to finding the optimal radius $\varepsilon^*$, we first consider the critical condition as:
\begin{eqnarray}\label{e:proof_xi_0}
&&\hspace{-0.75cm}\rho_I = \sum_{m=1}^\mu\Biggl[ \frac{ A_{\text{unit}}}{\pi} \cdot \\
&&\hspace{-0.75cm}\frac{ \mathcal{H}^2 }{\left(\! (\varepsilon^*\! \cos{(\xi_0 + \frac{2\pi m}{\mu})}\! -\! x)^2 \!+\! (\varepsilon^* \!\sin{(\xi_0 + \frac{2\pi m}{\mu})} \!-\! y)^2 \!+\! \mathcal{H}^2 \!\right)^2\!\!}\Biggr].\nonumber
\end{eqnarray}
In \eqref{e:proof_xi_0}, we can treat $x$ and $y$ as functions of $\xi_0$, and find $\xi_0$ that maximizes both $x$ and $y$. Taking the partial derivative of \eqref{e:proof_xi_0} w.r.t. $\xi_0$, we have
\begin{eqnarray}\label{e:partial_xi_0}
&&\hspace{-0.75cm} \sum_{m=1}^\mu \! \frac{ 1 }{\left[ \big[\varepsilon^* \cos{(\xi_0\! +\! \frac{2\pi m}{\mu})}\! -\! x\big]^2 \!\!\!+\!\! \big[\varepsilon^* \sin{(\xi_0 \!+\! \frac{2\pi m}{\mu})} \!-\! y\big]^2 \!\!\!+\!\! \mathcal{H}^2 \right]^3} \!\cdot \nonumber\\
&&\hspace{-0.75cm} \Biggl[ \big[\varepsilon^*\! \cos\!{\left(\! \xi_0\! +\! \frac{2\pi m}{\mu}\right)}\! -\! x\big] \!\!\cdot\!\! \left(-\varepsilon^*\! \sin\!{\left(\! \xi_0\! +\! \frac{2\pi m}{\mu}\right)}\! -\! \frac{\partial}{\partial \xi_0}x \right) \\
&&\hspace{-0.75cm} +\! \big[\varepsilon^*\! \sin\!{\left(\! \xi_0 \!+\! \frac{2\pi m}{\mu}\right)} \!-\! y\big] \!\!\cdot\!\! \left(\varepsilon^* \!\cos\!{\left(\! \xi_0 \!+\! \frac{2\pi m}{\mu}\right)} \!-\! \frac{\partial}{\partial \xi_0}y \right) \Biggr] =0.\nonumber
\end{eqnarray}
Since our goal is the extremum values of $x$ and $y$, we can set $\frac{\partial}{\partial \xi_0}x = \frac{\partial}{\partial \xi_0}y = 0$. Considering the critical points on the diagonal that are farthest from the center of the room, \eqref{e:partial_xi_0} can be refined as
\begin{eqnarray}\label{e:partial_xi_0_2}
\sum_{m=1}^\mu \! \frac{ -\sqrt{2}\varepsilon^*\sin{\left(\xi_0 + \frac{2 \pi m}{\mu}- \frac{\pi}{4}\right)} x }{\biggl[\varepsilon^{*2} \!+ \! \mathcal{H}^2 \! + \! 2 x^2 \!-\! 2 \sqrt{2}\varepsilon^*\!\sin{\!\left(\! \xi_0 \!+ \!\frac{2 \pi m}{\mu} \!+\! \frac{\pi}{4}\!\right)\!} x \biggr]^3}\! \!=\!0.
\end{eqnarray}
We define
\begin{eqnarray}\label{e:F_xi}
F(\xi_0) \!\triangleq \!\!\!\!\sum_{m=1}^\mu \!\!\! \frac{ -\sqrt{2}\varepsilon^*\sin{\left(\xi_0 + \frac{2 \pi m}{\mu}- \frac{\pi}{4}\right)} x }{\biggl[\varepsilon^{*2}\! \!+ \! \mathcal{H}^2 \! + \! 2 x^2 \!-\! 2 \!\sqrt{2}\varepsilon^*\!\sin{\!\left(\! \xi_0 \!+ \!\frac{2 \pi m}{\mu} \!+\! \frac{\pi}{4}\!\right)} x \!\biggr]^3\!}.
\end{eqnarray}

\begin{lem}\label{thm:symmetric}
$F(\xi_0)$ has a rotational symmetry about the $\xi_a = - \frac{2\pi}{\mu} + \frac{\pi}{4}$, that is,
\begin{eqnarray}\label{e:odd_F}
F(\xi_a -\xi_0) = - F(\xi_a +\xi_0).
\end{eqnarray}
\end{lem}

\begin{NewProof}
First, when $m=1$, we have
\begin{eqnarray}\label{e:odd_F_m=1}
&&\hspace{-0.4cm} F(\xi_a -\xi_0)|_{m=1} \nonumber\\
&&\hspace{-0.8cm} = \frac{ -\sqrt{2}\varepsilon^*\sin{\left(\xi_a -\xi_0 + \frac{2 \pi}{\mu}- \frac{\pi}{4}\right)} x }{\biggl[\varepsilon^{*2} + \mathcal{H}^2 + 2 x^2 - 2 \sqrt{2}\varepsilon^* \sin{ \left( \xi_a -\xi_0 + \frac{2 \pi }{\mu} + \frac{\pi}{4} \right)} x \biggr]^3 } \nonumber\\
&&\hspace{-0.85cm} \overset{(a)}{=} \frac{ \sqrt{2}\varepsilon^*\sin{ \xi_0} x }{\biggl[\varepsilon^{*2} + \mathcal{H}^2 + 2 x^2 - 2 \sqrt{2}\varepsilon^* \cos{ \xi_0} x \biggr]^3 } \nonumber\\
&&\hspace{-0.8cm} = - F(\xi_a +\xi_0)|_{m=1},
\end{eqnarray}
where (a) follows because $\xi_a = - \frac{2\pi}{\mu} + \frac{\pi}{4}$.

On the other hand, when $m=2,..., \mu$, we have
\begin{eqnarray}\label{e:odd_F_m=2}
&&\hspace{-0.4cm} F(\xi_a -\xi_0)|_{m \neq 1} \nonumber\\
&&\hspace{-0.8cm} = \frac{ -\sqrt{2}\varepsilon^*\sin{\left(\xi_a -\xi_0 + \frac{2 \pi m}{\mu}- \frac{\pi}{4}\right)} x }{\biggl[\varepsilon^{*2} + \mathcal{H}^2 + 2 x^2 - 2 \sqrt{2}\varepsilon^* \sin{ \left( \xi_a -\xi_0 + \frac{2 \pi m}{\mu} + \frac{\pi}{4} \right)} x \biggr]^3 } \nonumber\\
&&\hspace{-0.8cm} = \frac{ -\sqrt{2}\varepsilon^*\sin{\left( -\xi_0 + \frac{2 \pi (m-1)}{\mu}\right)} x }{\biggl[\varepsilon^{*2} + \mathcal{H}^2 + 2 x^2 - 2 \sqrt{2}\varepsilon^* \cos{ \left( -\xi_0 + \frac{2 \pi (m-1)}{\mu} \right)} x \biggr]^3 } \nonumber\\
&&\hspace{-0.8cm} = - F(\xi_a +\xi_0)|_{\mu + 2 - m},
\end{eqnarray}
proving Lemma \ref{thm:symmetric}.
\end{NewProof}

Since $F(\xi_a -\xi_0) = - F(\xi_a +\xi_0)$ and $F(\xi_a) = - F(\xi_a )$, the optimal solution to \eqref{e:partial_xi_0_2} is given by $\xi_0^* = - \frac{2\pi }{\mu} + \frac{\pi}{4}$. Theorem \ref{thm:dist} is proved.

\begin{figure}[!t] 
\centering{\includegraphics[width=0.8\linewidth]{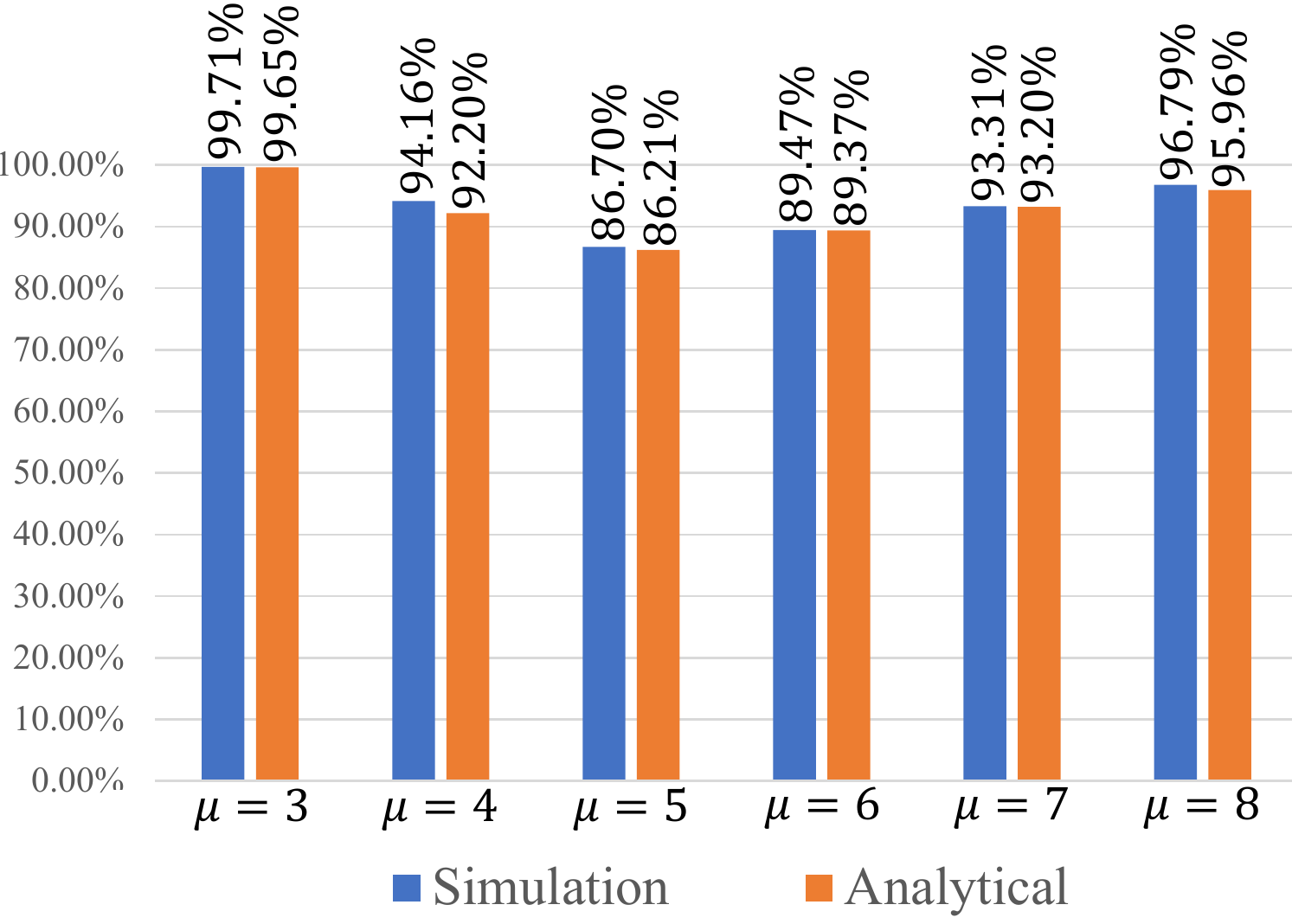} }
\caption{The comparison between the simulated optimal solution and the analytical solution regarding the proportion of area exceeding the threshold.}
\label{f:diag_approx}
\end{figure}

\begin{figure}[!t] 
\centering{\includegraphics[width=0.6\linewidth]{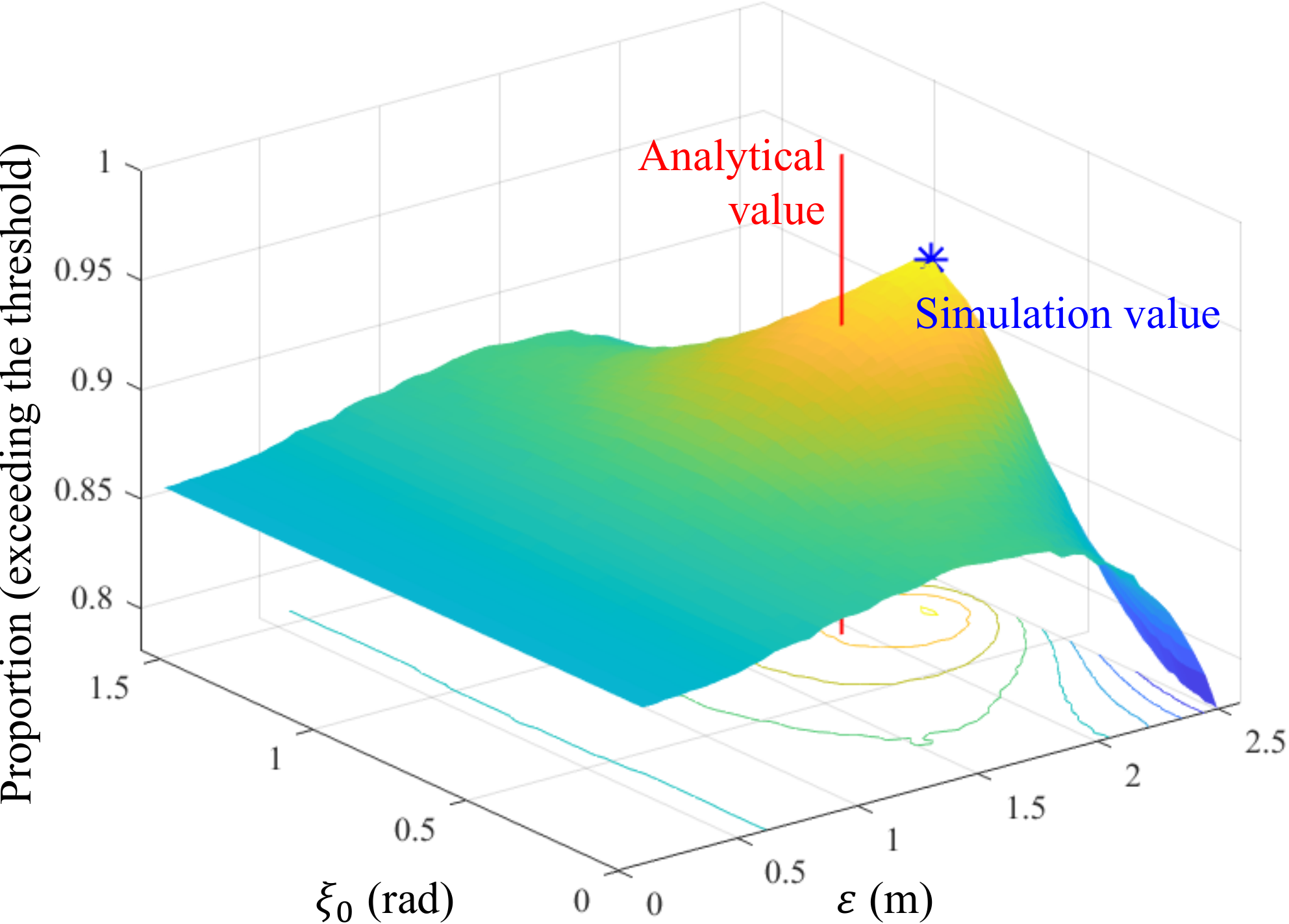} }
\caption{The proportion of the area exceeding the threshold under various $\varepsilon$ and $\xi_0$. The vertical line indicates the analytical value obtained from Theorem~\ref{thm:dist}.}
\label{f:surf_prop}
\end{figure}

To validate Theorem \ref{thm:dist}, Fig.~\ref{f:diag_approx} compares the simulated optimal solution from \eqref{e:optimization_h} with the analytical solution in Theorem \ref{thm:dist}. Details of the simulation setup are in Tab.~\ref{t-para} and will be further discussed in Sec.~\ref{sec:V}. The results show that variations in the area proportion exceeding the threshold are negligible across different $\mu$ values. For instance, a 0.5 m$^2$ area difference for $\mu=4$ case implies the simulated optimal solution adds an annular region about 51 mm wide.
Furthermore, in Fig.~\ref{f:surf_prop}, we present the simulation results obtained by traversing all possible values of $\varepsilon$ and $\xi_0$ when $\mu=4$. We found that the angles in the analytical solution are the same as those in the simulated optimal solution, while a similar distribution radius.
As shown, the analytical solution given in Theorem \ref{thm:dist} is a good approximation to the optimal solution of \eqref{e:optimization_h}.

\section{Optical beamforming}\label{sec:IV}
In the first phase of O-ISAC, the O-APs broadcast control messages to all devices and sense the devices' state with the reflected light. Building on the analysis presented in Sec.~\ref{sec:III}, we recognize the pivotal role of superimposed light intensity on the target, or the equivalent channel gain \eqref{e:equivalent_channel}, in enhancing system reliability. Consequently, during Phase 2, to enhance communication and sensing performance, we position all O-APs to directly face the target device—moving away from their initial vertical alignment—and employ a TDMA strategy for device service. However, despite these adjustments, the signal's intensity received by the device remains constrained due to the inherent divergence of directionless light.
In RF communication, an efficient scheme to address such a problem is beamforming, which uses antenna arrays to direct the signal toward specific angles.
In optical communication, however, beamforming cannot be realized due to the uncontrollable phase of light.

To achieve spatial selectivity, this paper puts forth the concept of optical beamforming for O-ISAC, leveraging the collimating lens.
This approach is predicated on the assumption that we can align the O-APs with the target device using a three-dimensional turntable. As a result, our focus shifts towards designing a collimating lens, a type of curved optical lens that aligns light in parallel, effectively mitigating the continuous diffusion of light.
Inspired by this, we design collimating lenses to modify the radiation pattern $R(\varphi_m)$ of the light source and direct all emitted light to the target receiver, achieving the effect of optical beamforming.

\begin{figure}[!t]
\centering
\includegraphics[width=0.7\linewidth]{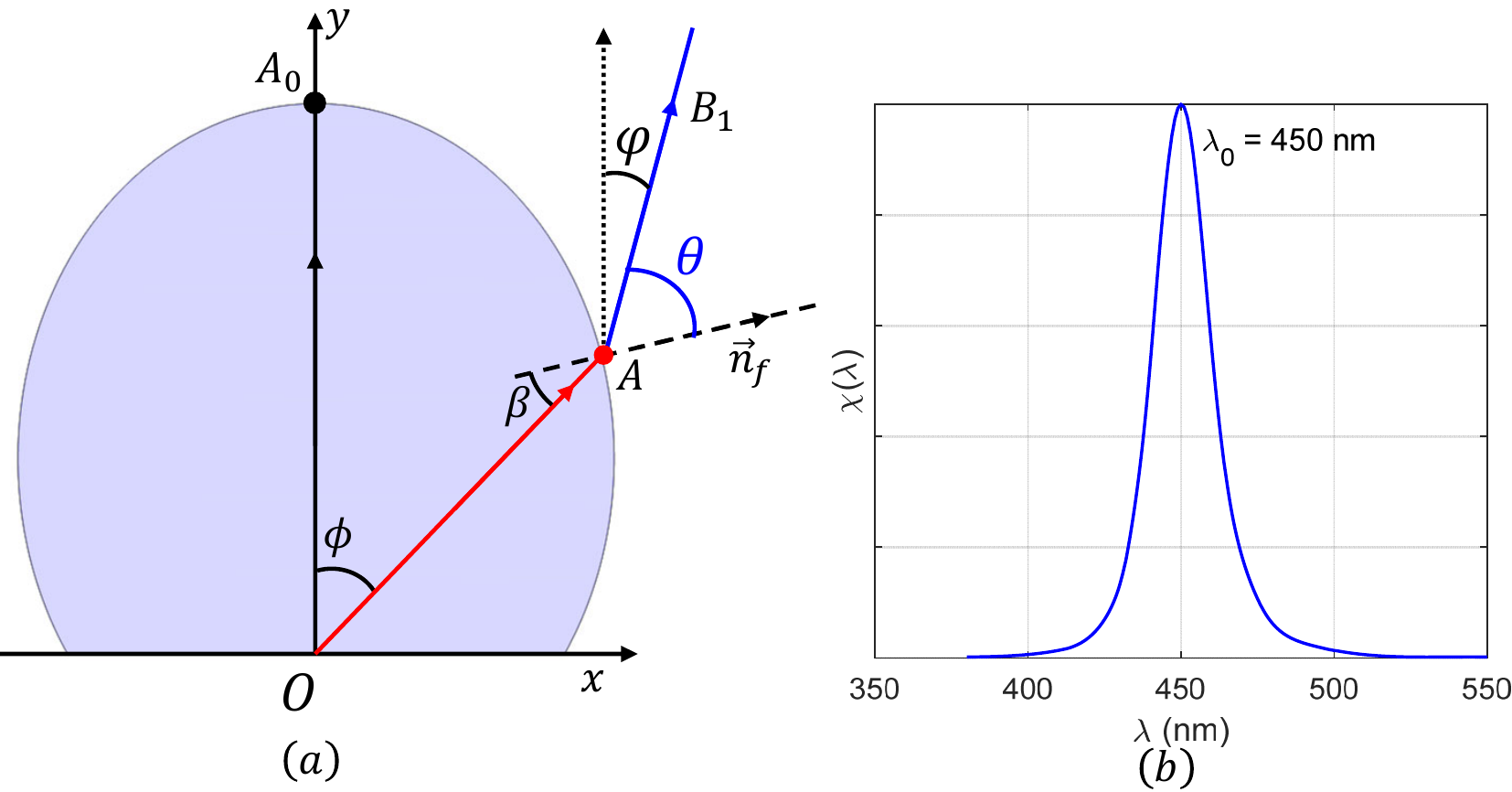}
\caption{(a) The cross-section of a collimating lens. 
(b) The relative spectrum density of a commercial LED. 
  }
\label{f:Obf}
\end{figure}

The radiation pattern of an optical source can be changed when a collimating lens is introduced. As shown in Fig.~\ref{f:Obf}(a), a light ray with an AoE of $\phi$ is emitted from the light source and directly enters the lens. The ray intersects with the lens surface on point $A$ and undergoes refraction as it enters the air. 
We denote by $\overrightarrow{n}_f$ the normal vector of the lens at point A.
The angle at which the ray enters the lens is called the incident angle $\beta$, while the angle at which it exits the lens surface is referred to as the exit angle $\theta$. 
The angle between the refracted light ray after leaving the lens surface and the $y$-axis is the AoD defined earlier in Sec.~\ref{sec:II}.
Geometrically, the relationship between these angles can be expressed as $\phi = \varphi - \beta  + \theta $.

A challenge here is that practical optical sources cannot produce light of a single frequency \cite{saleh2019fundamentals}, which is commonly referred to as the frequency dispersion effect.
As an example, Fig.~\ref{f:Obf}(b) depicts the relative spectrum density $\chi(\lambda)$ of the light emitted by a commercial LED as a function of wavelength. In addition to the peak wavelength $\lambda_0=450$ nm, the emitted light also contains other wavelengths.
This implies that,
\begin{itemize}
    \item For the light beam with a given AoE $\phi$, different wavelength components $\lambda$ in the beam undergo different refraction through the lens. In this case, $\theta$ and $\varphi$ can be written as functions of $\lambda$, i.e., $\theta(\lambda|\phi)$ and $\varphi(\lambda|\phi)$.
    \item The light beam with a given AoD $\varphi$, on the other hand, consists of a cluster of light rays with different AoE $\phi$ and wavelength $\lambda$. In this case, $\phi$, $\beta$ and $\theta$ can be written as functions of $\lambda$, i.e., $\phi(\lambda|\varphi)$, $\beta(\lambda|\varphi)$ and $\theta(\lambda|\varphi)$.
\end{itemize}

Considering a specific target device located at $\bm{p}_D = (x_D, y_D, z_D)$. Suppose the length and width of the target are $\mathcal{L}_t$ and $\mathcal{W}_t$, respectively. The radiation pattern optimization problem can be formulated as maximizing the sum of light intensities falling on the target device, yielding
\begin{subequations}\begin{align}\text {(P2):}~&\max_{\{R(\varphi_m)\}} \int_{y_D- \frac{\mathcal{L}_t}{2}}^{y_D+ \frac{\mathcal{L}_t}{2}}{\int_{x_D - \frac{\mathcal{W}_t}{2}}^{x_D + \frac{\mathcal{W}_t}{2}}{ \sum\limits_{m} {I_{m,k} ^ { \text{rx} } }  dx }dy },\label{e:P2}
\\ {\text {s.t.}~} 
    &I_{m,k}^{\text{rx}} = R\big(\varphi_m \big)\cdot\frac{ A_{\text{unit}} \cos{\psi_m} }{  d_{m,k}^2} ,\label{e:P2I}
    \\&  R(\varphi_m )= \int{\frac{ 1 }{ \pi } \cos \big( \phi_{ m }\left( \lambda|\varphi_m \right) \big) \chi(\lambda) d\lambda },\label{e:P2R}
  \\ d_{m,k} =& \sqrt{(\varepsilon^* \cos{ \xi_m^*} \!-\! x_{D})^2 \!+\! (\varepsilon^* \sin{ \xi_m^*} \!-\! y_{D})^2 \!+\! \mathcal{H}^2}, \label{e:P2d}
  \\& \sin \beta \cdot n(\lambda)  = \sin \big(\theta(\lambda|\varphi_m)\big) \cdot 1 , \label{e:P2ref}
  \\
  &\phi_m(\lambda|\varphi_m) = \varphi_m - \beta(\lambda|\varphi_m) + \theta(\lambda|\varphi_m), \label{e:P2phi}
     \end{align}  \end{subequations} 
where \eqref{e:P2ref} is the law of refraction and $n(\lambda)\propto 1/\lambda$ represents the refractive index of the lens.

With optical beamforming, the light emitted by O-APs is intentionally directed towards the target device, enhancing both the optical communication rate and the sensing accuracy, thanks to the much stronger received light intensity.
Theorem \ref{thm:ob} below summarizes our main result in this section. 
\begin{thm}\label{thm:ob}
In the second phase of O-ISAC, with optical beamforming, the optimal radiation pattern of a Lambertian optical source that maximizes the received light intensity, i.e., the optimal solution to (P2), can be approximated by
\begin{equation}\label{e:R*}
R^*\big( \varphi_m \big)= \int{\frac{ 1 }{ \pi } \cos\left( \frac{\big(n(\lambda)-1 \big)\lambda}{n(\lambda)(\lambda -\lambda_0)} \varphi_m  \right) \chi(\lambda) d\lambda }.
\end{equation}
\end{thm}

In RF communication, antenna arrays with phase shifter networks enhance directivity and enable beam scanning. Yet, applying these methods to the THz band is complex and costly, especially for non-coherent sources like LEDs. Therefore, lens-based beamforming presents a practical alternative \cite{abbasi2019constant}, and we prove Theorem \ref{thm:ob} by a geometric approach. In \eqref{e:P2}, the parameters $x_D$, $y_D$, $\mathcal{W}_t$, and $\mathcal{L}_t$ depend on the position and size of the target device. Since in the second phase, we have aligned all the O-APs towards the target device, the optimization objective in \eqref{e:P2} is equivalent to maximizing the light intensity $I^\text{rx}_{m,k}$ in \eqref{e:P2I} at the target device. 
In other words, our objective is to identify the highest achievable radiation pattern $R(\varphi_m)$ for the O-APs when $\varphi_m=0$.
Since the radiation of the optical source is independent, we can optimize the radiation pattern of the LEDs independently. Without loss of generality, we consider the $m$-th LED in the following. The subscript $m$ will be omitted for simplicity.

According to \eqref{e:P2R}, ideally, we desire that for any AoE $\phi$, the light ray after collimation has an AoD $\varphi(\lambda|\phi)=0$, $\forall \lambda$. This maximizes $R(\varphi=0)$.
However, one problem is that the lens can only be designed to perfectly collimate one wavelength.
In this context, $R(\varphi=0)$ is optimized if and only if $\varphi(\lambda=\lambda_0|\phi)=0$, $\forall \phi$.

\begin{lem}[Profile of the lens surface]\label{prop:lens}
The optimal normal vector of the collimating lens that satisfies $\varphi(\lambda_0|\phi)=0$, $\forall \phi$, at any point on the lens surface is given by
\begin{eqnarray}
    &&\hspace{-1.5cm} \overrightarrow{n }^*_f  =\big(n(\lambda_0)\sin{\phi} , n(\lambda_0)\cos{\phi} -  1\big).
\end{eqnarray}
\end{lem}

\begin{figure}[!t]
\centering
\includegraphics[width=0.7\linewidth]{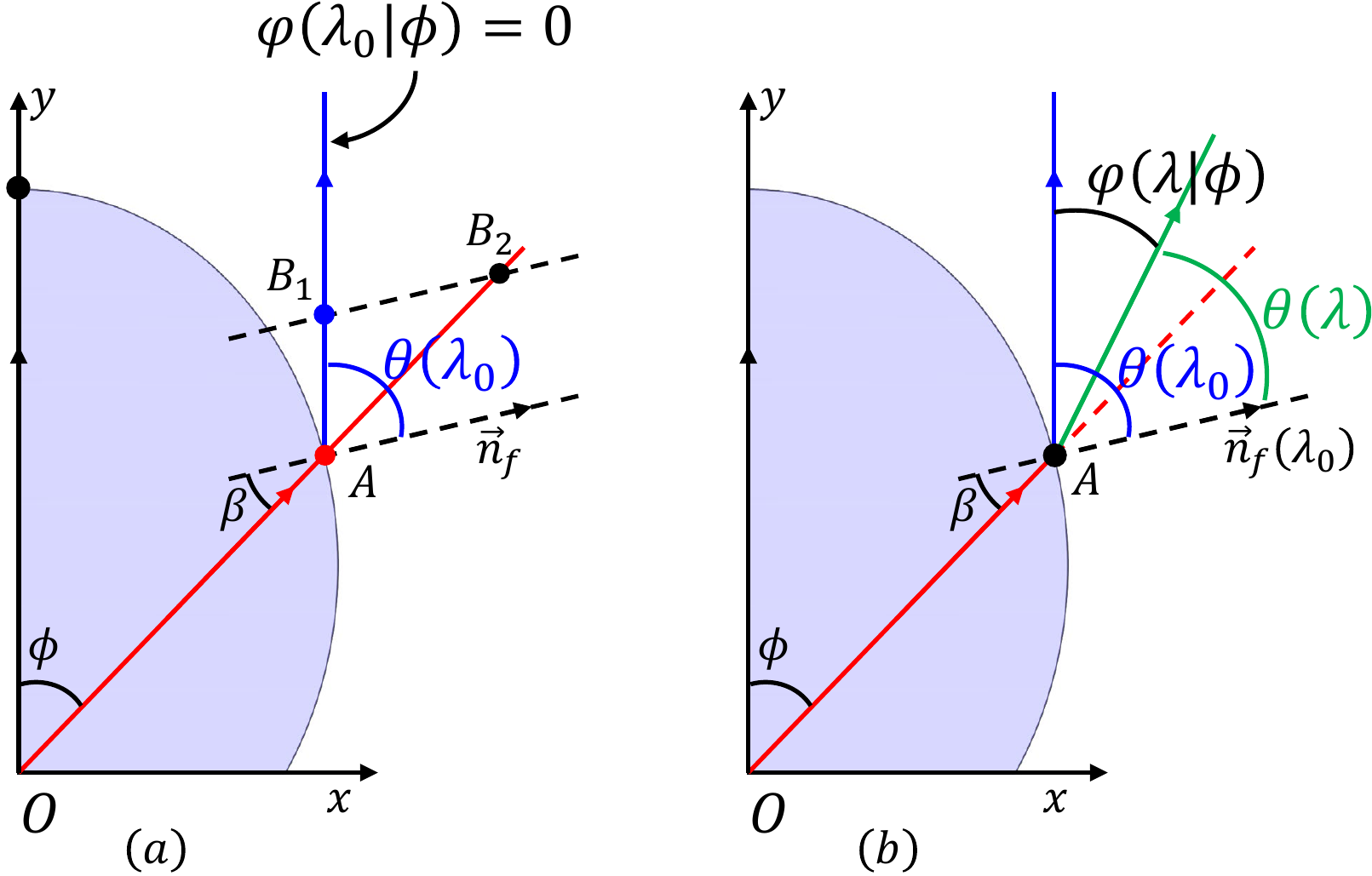}
\caption{The cross-section of a collimating lens. }
\label{f:lens}
\end{figure}

\begin{NewProof}
As shown in Fig.~\ref{f:Obf}(a), the direction of a light ray $\overrightarrow{\text{OA}}$ can be represented as
$\frac{\overrightarrow{ \text{OA} }} { | \overrightarrow{ \text{OA}} | } = (\sin\phi,\cos\phi)$,
the direction of the outgoing light ray $\overrightarrow{\text{AB}_1} (\lambda)$ can be represented as 
$\frac{\overrightarrow{ \text{AB}_1  } (\lambda)} { | \overrightarrow{ \text{AB}_1 } (\lambda)| } = \big(\sin\varphi(\lambda|\phi),\cos\varphi(\lambda|\phi)\big),$
and our goal is
\begin{eqnarray} \label{e:AB1}
    \frac{\overrightarrow{ \text{AB}_1 } (\lambda_0)} { | \overrightarrow{ \text{AB}_1}(\lambda_0) | } = \Big(\sin\big(\varphi(\lambda_0|\phi)\big), \cos\big(\varphi(\lambda_0|\phi)\big)\Big) = (0,1).
\end{eqnarray}
An illustration of \eqref{e:AB1} is given in Fig.~\ref{f:lens}(a). At point B$_1$, we draw a line parallel to the normal $\overrightarrow{n}_f$, which intersects the extension of $\overrightarrow{ \text{OA} }$ at point B$_2$. Using the cosine rule, we have
\begin{eqnarray} \label{e:cosine}
    | \overrightarrow{ \text{AB}_1}(\lambda_0) |\sin{\theta} = | \overrightarrow{ \text{AB}_2} | \sin{\beta}.
\end{eqnarray}
Combining \eqref{e:P2ref} and \eqref{e:cosine}, we get
\begin{eqnarray} \label{e:length}
    | \overrightarrow{ \text{AB}_2} | = n(\lambda_0) | \overrightarrow{ \text{AB}_1}(\lambda_0) | .
\end{eqnarray}
Therefore, based on the vector triangle AB$_1$B$_2$, we can determine the optimal normal vector direction as 
\begin{eqnarray} \label{e:nf}
    &&\hspace{-1 cm}\overrightarrow{n}_f^* \propto 
    \frac{1} { | \overrightarrow{ \text{AB}_1}(\lambda_0) | }\left[\overrightarrow{ \text{AB}_2} -\overrightarrow{ \text{AB}_1}(\lambda_0)  \right]\nonumber\\
    &&\hspace{-0.4 cm} \overset{(a)}{=} \frac{1} { | \overrightarrow{ \text{AB}_1}(\lambda_0) | }\left[ \frac{\overrightarrow{ \text{AB}_2 }} { | \overrightarrow{ \text{AB}_2} | } \cdot n(\lambda_0) | \overrightarrow{ \text{AB}_1}(\lambda_0) | - \overrightarrow{ \text{AB}_1}(\lambda_0) \right]\nonumber\\
    &&\hspace{-0.4 cm} \overset{(b)}{=} \big(n(\lambda_0)\sin{\phi} , n(\lambda_0)\cos{\phi} -  1\big),
\end{eqnarray}
where (a) follows from \eqref{e:length}, (b) follows from $\frac{\overrightarrow{ \text{AB}_2 }} { | \overrightarrow{ \text{AB}_2} | }  = \frac{\overrightarrow{ \text{OA} }} { | \overrightarrow{ \text{OA}} | } $ and \eqref{e:AB1}.
This proves Lemma \ref{prop:lens}.
\end{NewProof}

The collimating lens given in Lemma \ref{prop:lens} effectively manipulates the peak wavelength component $\lambda_0$ emitted at various AoE, directing them towards the target user. On the other hand, other wavelength components cannot be fully collimated and still suffer from a divergence angle, i.e., AoD $\varphi(\lambda|\phi) \neq 0$, $\forall \lambda\neq\lambda_0$, as visualized in Fig.~\ref{f:lens}(b).

\begin{lem}[AoD]\label{thm:2}
Consider an optical source and denote the peak wavelength by $\lambda_0$.
After passing through the collimating lens characterized in Lemma \ref{prop:lens}, the AoD of a wavelength component $\lambda$ can be approximated by
\begin{equation}\label{e:theta_out}
\varphi(\lambda|\phi) \approx \frac{n(\lambda)}{n(\lambda)-1}\frac{\lambda-\lambda_0}{\lambda} \phi.
\end{equation}
\end{lem}

\begin{NewProof}
 See Appendix \ref{Appendix_exit_angle}.
\end{NewProof}

\begin{figure}[!t] 
\centering\includegraphics[width=.8\linewidth]{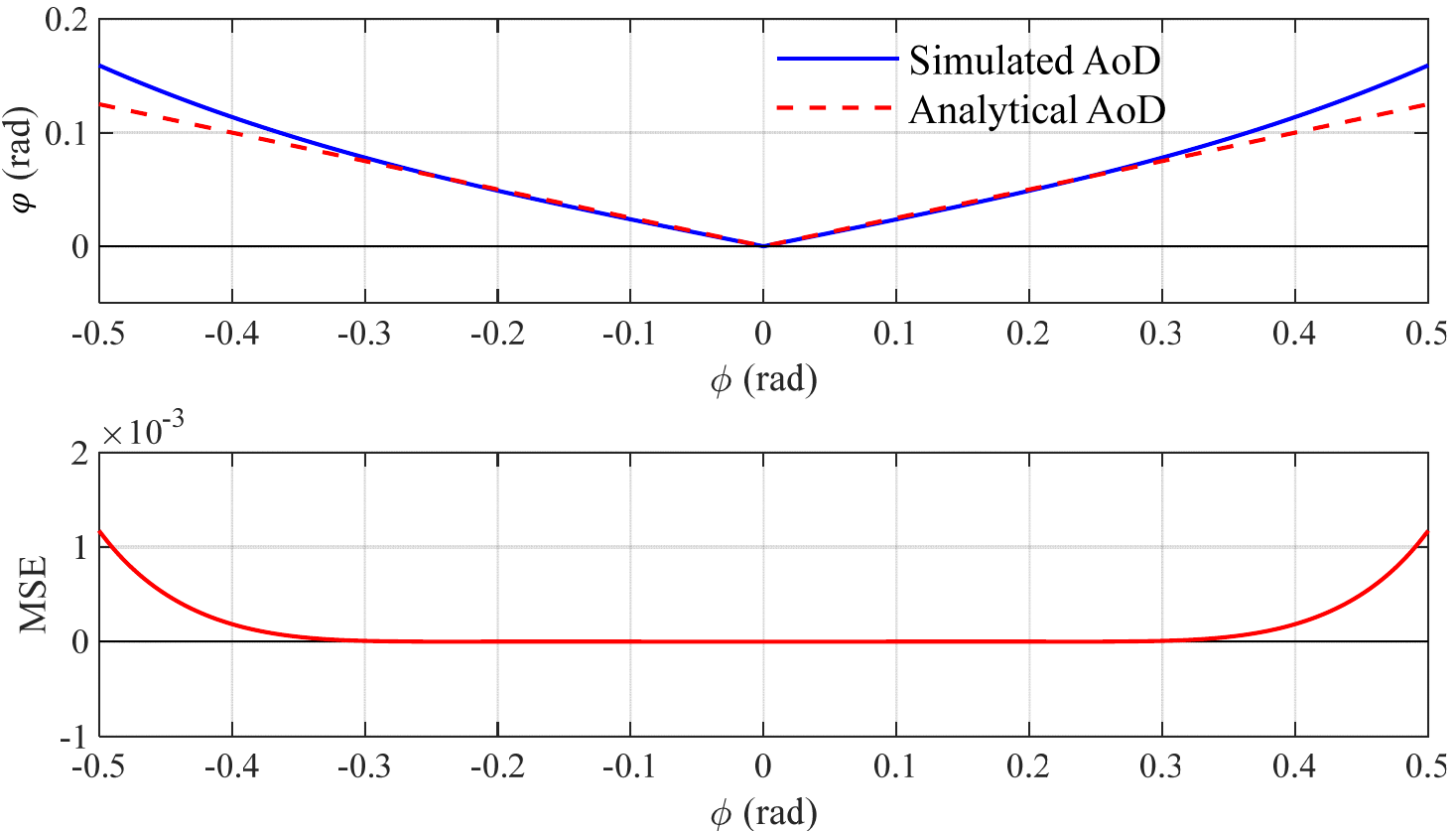} 
\caption{Comparison between simulated AoD and analytical AoD of the divergence angle after collimation, where $\lambda_0=450$ nm, $\lambda= 420$ nm, and $n(\lambda_0)=1.4$.}
\label{f:approx}
\end{figure}

To validate Lemma \ref{thm:2}, we compare the simulated and analytical values of $\varphi$ in Fig. \ref{f:approx}, where $\lambda_0 =450$ nm, $\lambda = 420$ nm, and $n(\lambda_0)=1.4$. As shown, \eqref{e:theta_out} is a decent approximation of $\varphi$, especially when the AoE $\phi$ is small.
From \eqref{e:theta_out}, we have
\begin{equation}\label{e:phi_varphi}
\phi( \lambda| \varphi) \approx \frac{\big(n(\lambda)-1 \big)\lambda}{n(\lambda)(\lambda -\lambda_0)} \varphi.
\end{equation}
Substituting \eqref{e:phi_varphi} into \eqref{e:P2R}, we obtain the optimal radiation pattern $R^*(\varphi)$ in \eqref{e:R*}, proving Theorem \ref{thm:ob}.

\section{Numerical and Simulation Results}\label{sec:V}

\begin{figure}
    \centering
    \begin{subfigure}[!t]{0.48\linewidth}
        \centering
        \includegraphics[width=\linewidth]{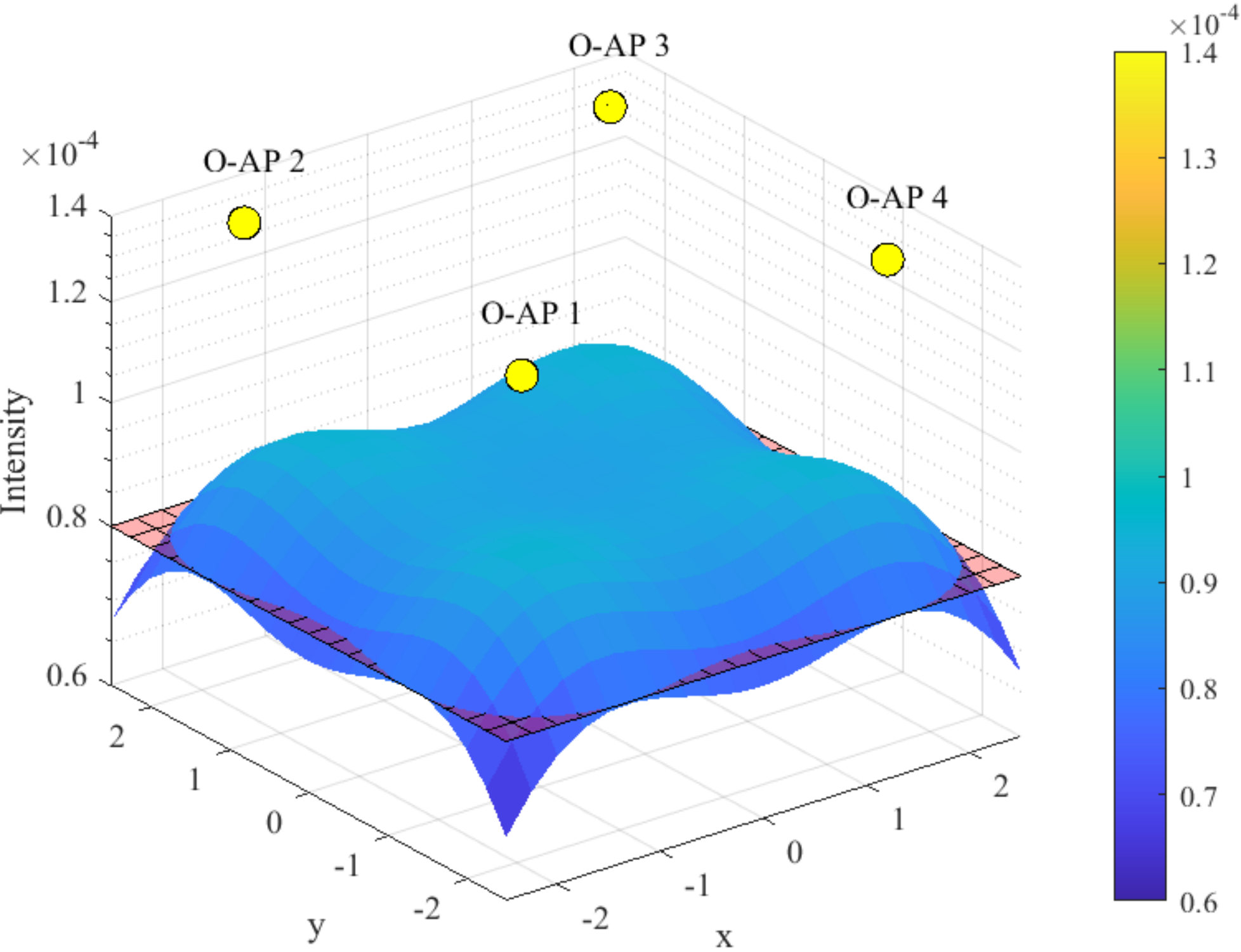}
        \caption{Light intensity (uniformity)}
        \includegraphics[width=\linewidth]{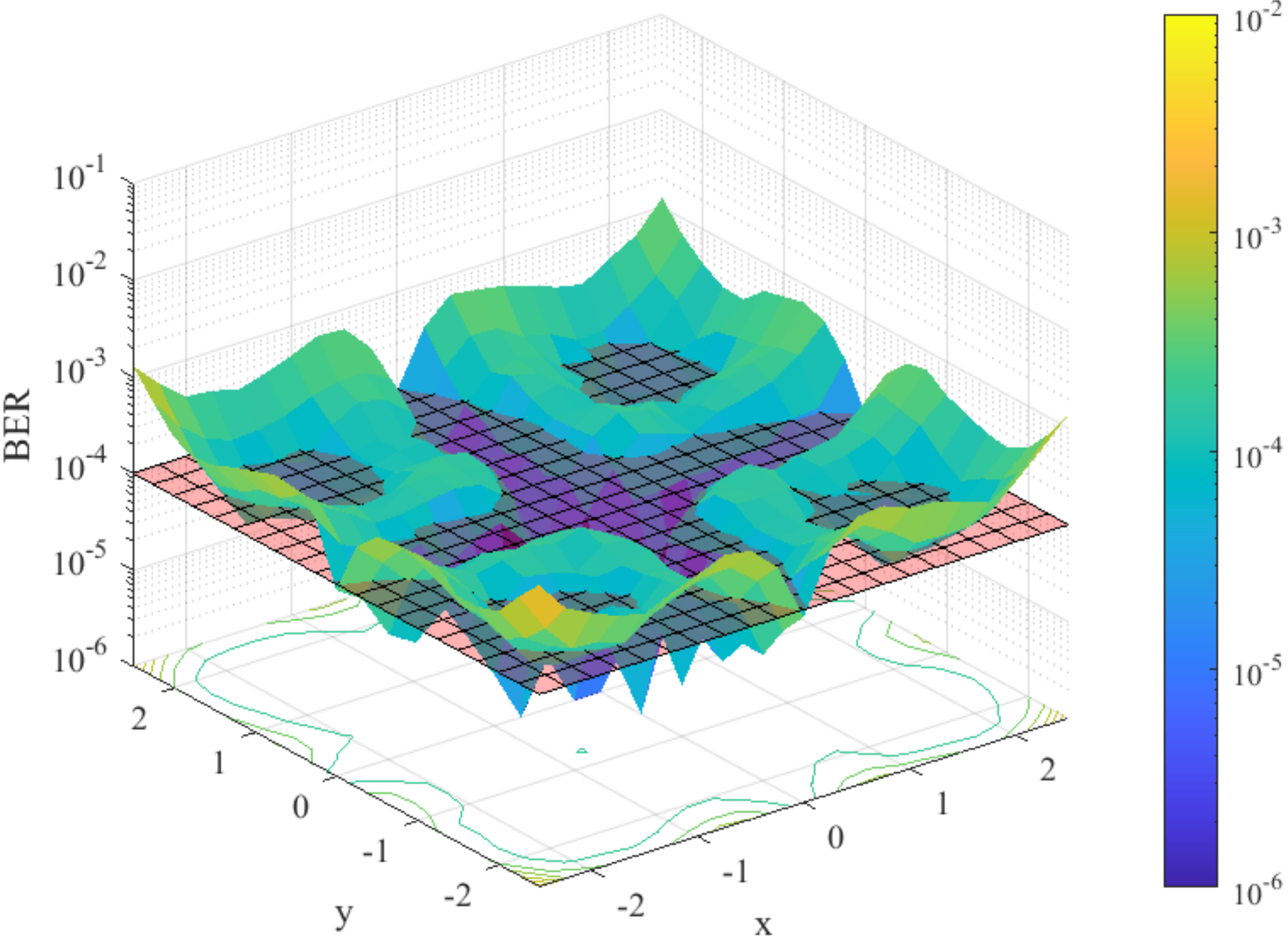}
        \caption{BER (uniformity)}
        \includegraphics[width=\linewidth]{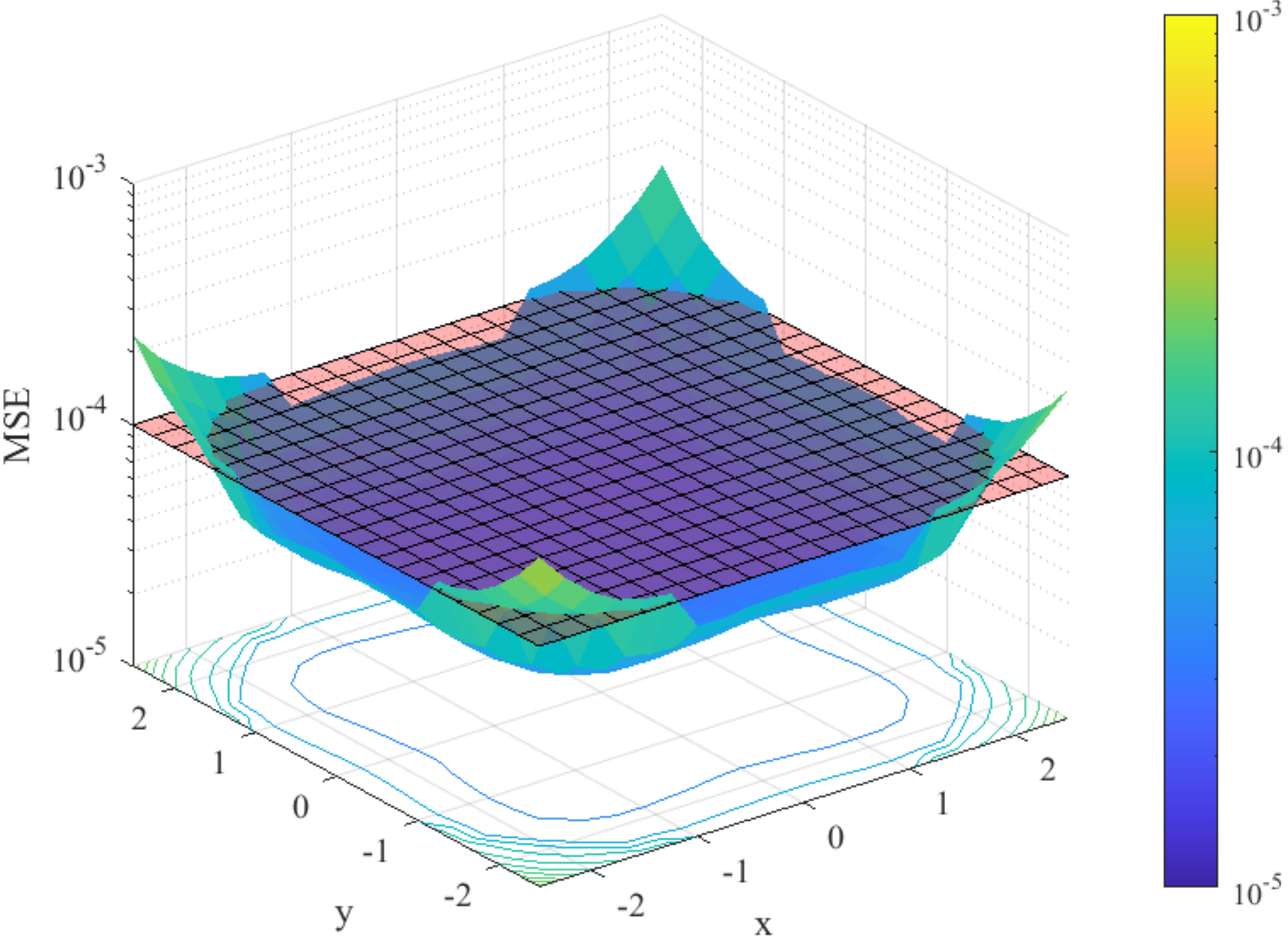}
        \caption{MSE (uniformity)}        
    \end{subfigure}
    \begin{subfigure}[!t]{0.48\linewidth}
        \centering
        \includegraphics[width=\linewidth]{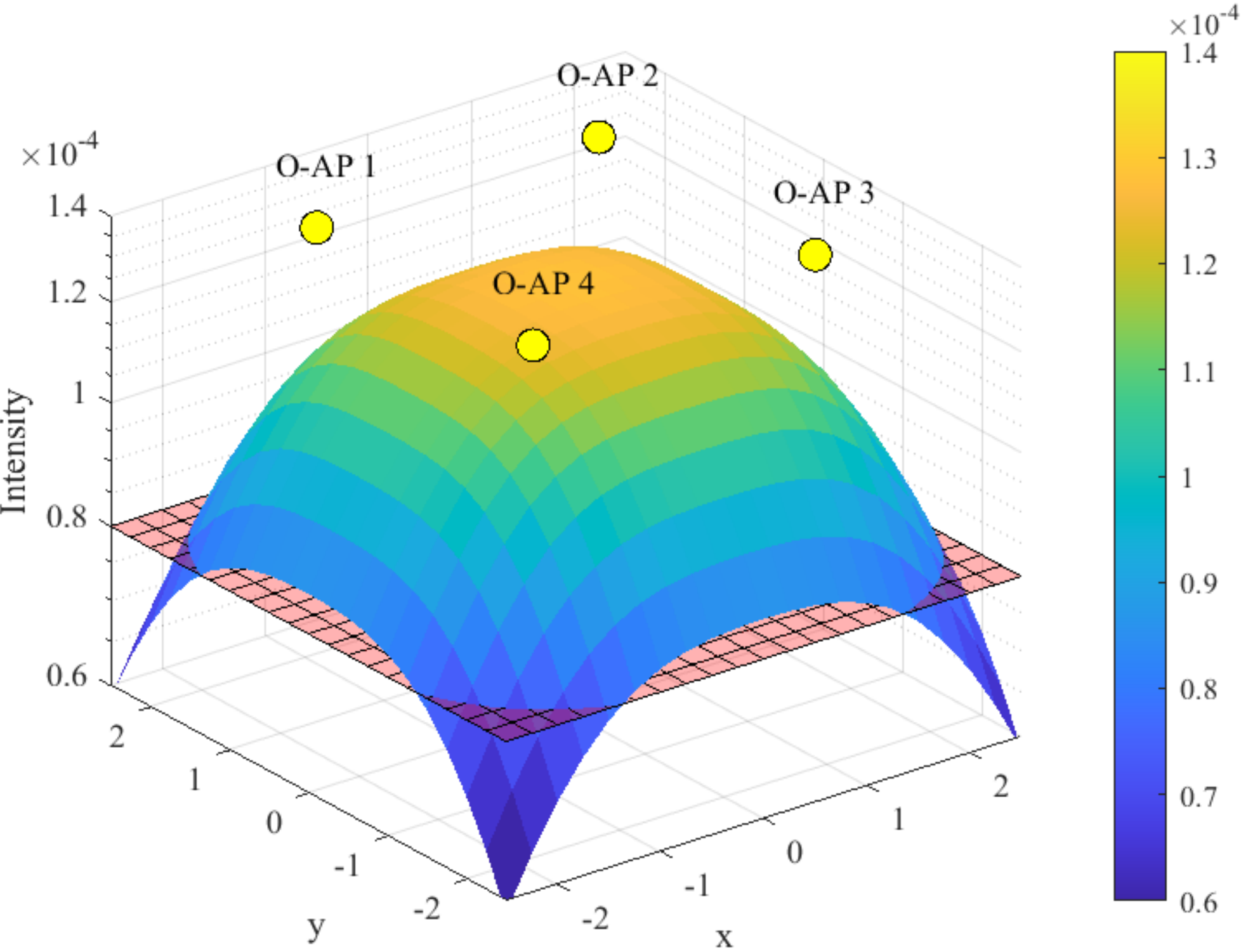}
        \caption{Light intensity (threshold)}
        \includegraphics[width=\linewidth]{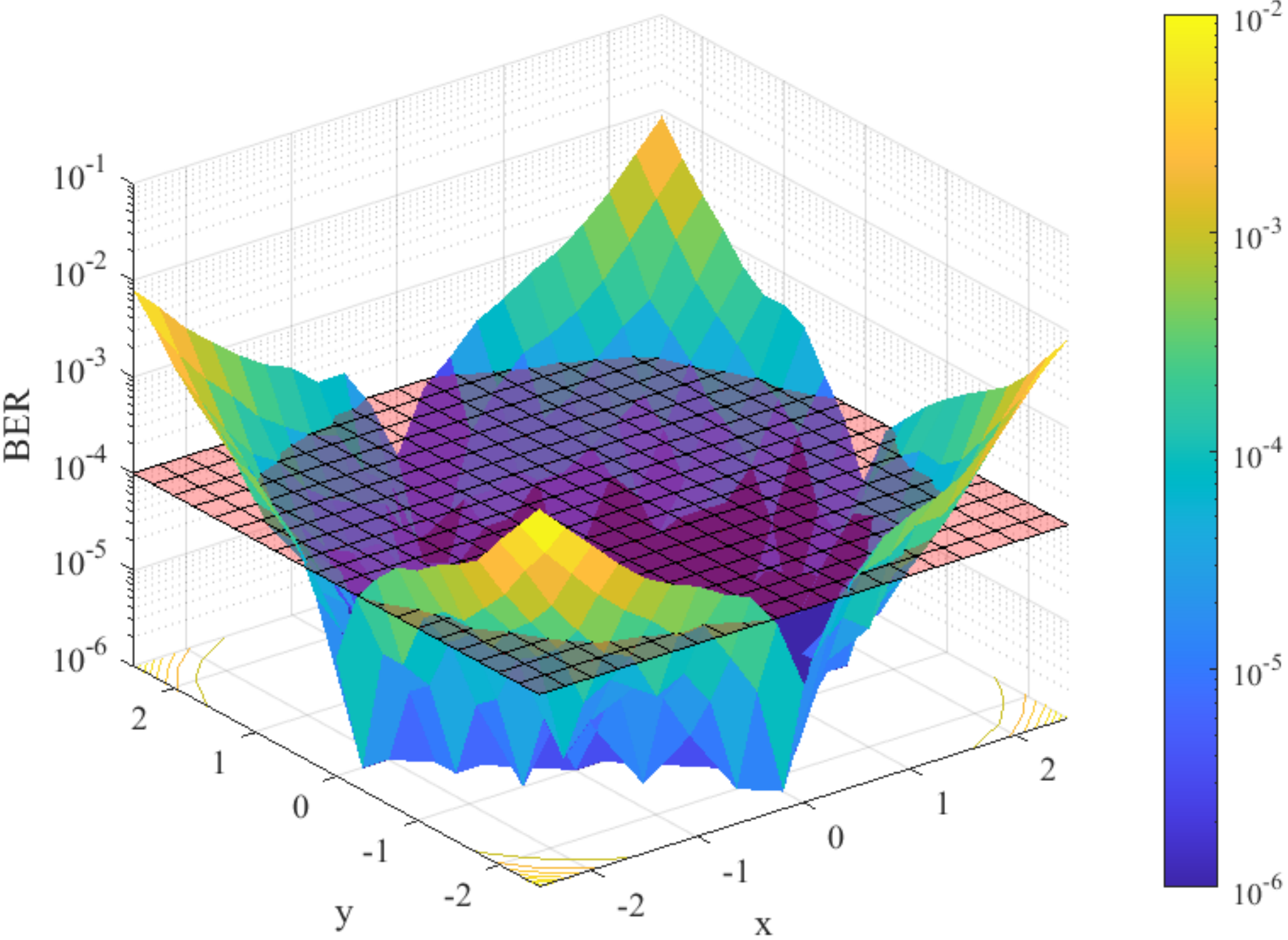}
        \caption{BER (threshold)}
        \includegraphics[width=\linewidth]{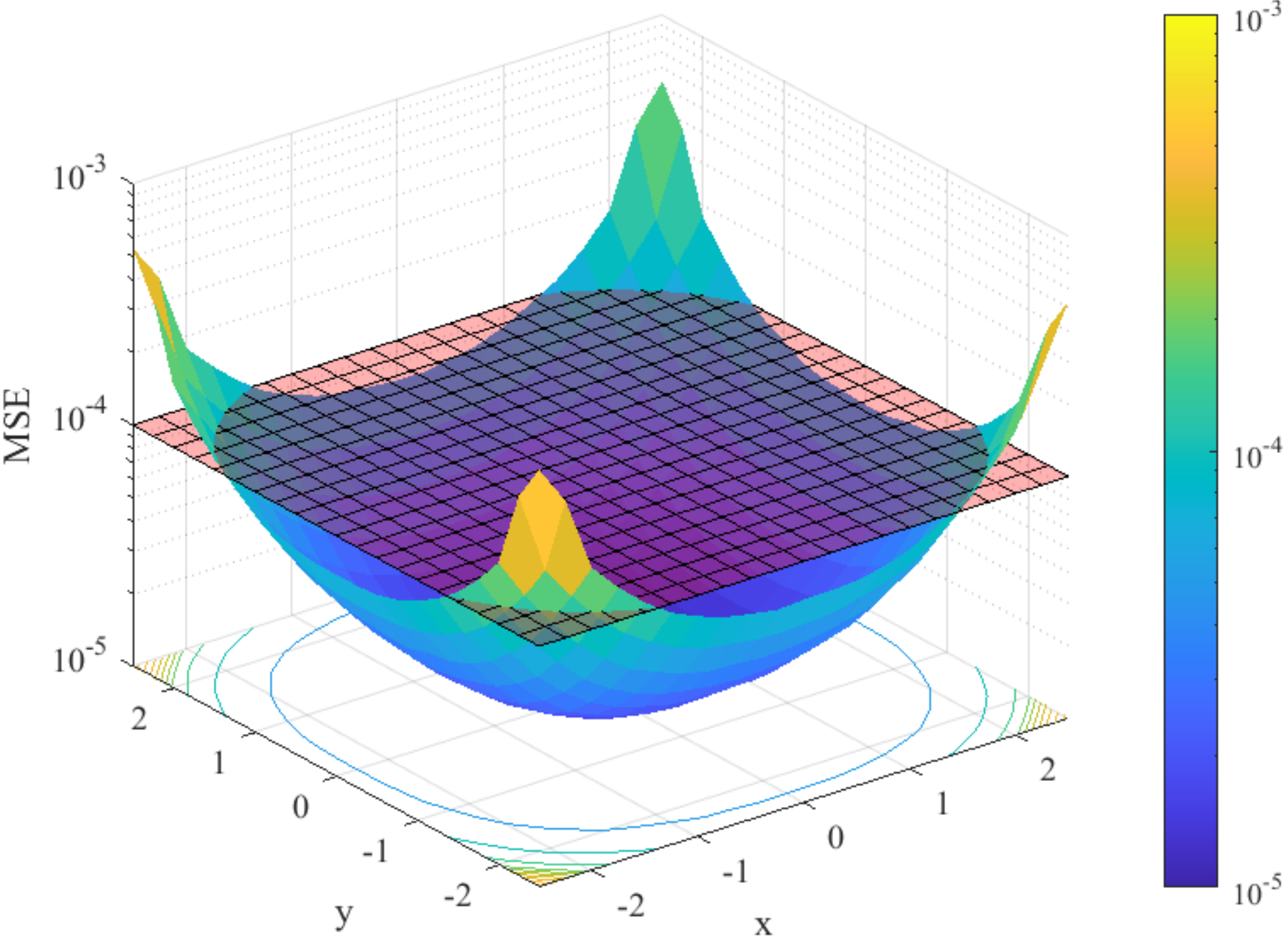}
        \caption{MSE (threshold)}
    \end{subfigure}
    \caption{The light intensity, BER, and MSE distributions under two optimization methods:
(a, d) the distribution of light intensity; (b, e)  the distribution of BER performance; (c, f) the distribution of MSE performance.}
    \label{f:coverage}
\end{figure}

This section assesses the effectiveness of the proposed O-ISAC system through numerical and simulation results. Specifically, we will conduct evaluations for both the directionless and directional O-ISAC, comparing them against a setup where sensing and communication operate independently, each with half the power.
To provide a clearer picture, the simulation setup and parameter settings are summarized in Tab.~\ref{t-para}.

\begin{table}[!t]
\centering
\caption{Parameter settings.}
\begin{tabular}{c|c|c}
\hline
\hline
\textbf{Parameters} & \textbf{Description} &\textbf{Value}\\
\hline
Environment
& Room dimension
& $5 \text{m}\times5 \text{m}\times3 \text{m}$\\\hline
\multirow{3}{*}{\makecell[c]{ Signal\\Structure}}
& No. of bits  & $2\times 10^5$\\ \cline{2-3}
& No. of subcarrier  & 32\\ \cline{2-3}
& Modulation scheme & \makecell[c]{BPSK-OFDM (phase 1)\\16QAM-OFDM (phase 2)}\\\hline
\multirow{6}{*}{\makecell[c]{Source\\Parameters}}
& No. of LED  & 4\\ \cline{2-3}
& Source Co-ordinates &$(\varepsilon \cos{\xi_m}, \varepsilon \sin{\xi_m}, \mathcal{H})$\\ \cline{2-3}
&\makecell[c]{Semi-half angle of LED\\ ($\Phi_{1/2}$)} & $\pi/3$
\\ \cline{2-3}
& focal length  & $0.05$ m\\ \hline
\multirow{5}{*}{\makecell[c]{device\\Parameters}}
& No. of PD per device  & 4\\ \cline{2-3}
& FOV ($\Psi_{FOV}$)  & $\pi/3$ \\ \cline{2-3}
& Active area of PD ($A_{PD}$)  & 1 mm$^2$\\ \cline{2-3}
& Size of lens  & 1 inch\\ \cline{2-3}
& Reflection coefficient ($\rho_j$)  & 0.8\\\hline \hline
\end{tabular}
\label{t-para}
\end{table}

Our initial focus centers around the optimal light source distribution in the first phase.
We will compare the traditional criterion in \eqref{e:Intensity_MSE}, which minimizes the MSE of the light intensity, with the proposed criterion in \eqref{e:optimization_h}, which maximizes the area in which the light intensity surpasses a threshold.
For the proposed criterion, we use the approximated optimal solution given in Theorem \ref{thm:dist}, and the threshold of the received light intensity is set to $\rho_I = 0.8\times10^{-4}$. This threshold was derived by back-calculating from benchmark values for BER and MSE, which we will introduce shortly.
Fig.~\ref{f:coverage} presents the achieved performances of the two optimization criteria, where (a) and (d) are the light intensity distribution across the entire room, (b) and (e) are the distribution of the achieved BER, and (c) and (f) are the distribution of the achieved MSE. Note that although Fig.~\ref{f:coverage} focuses on comparing the estimation performance at a single height for a straightforward comparison, our paper's theoretical framework is well-suited for
scenarios with multiple users at different heights.

\begin{figure}[!t] 
\centering{\includegraphics[width=0.7\linewidth]{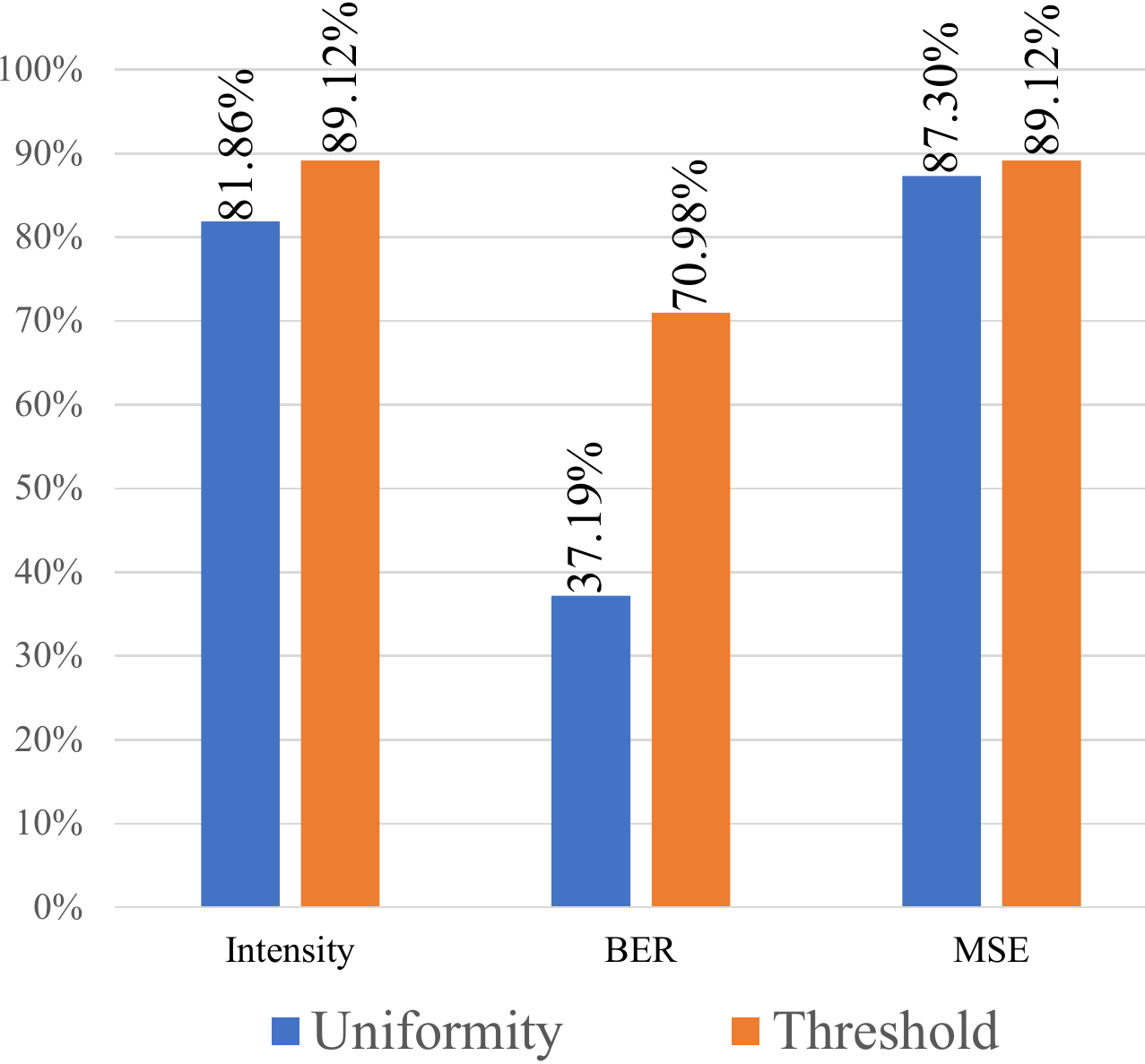} }
\caption{The proportion of the area achieving performance surpassing the threshold using the two optimization methods.}
\label{f:bar}
\end{figure}

\begin{figure}[!t] 
\centering\includegraphics[width=.9\linewidth]{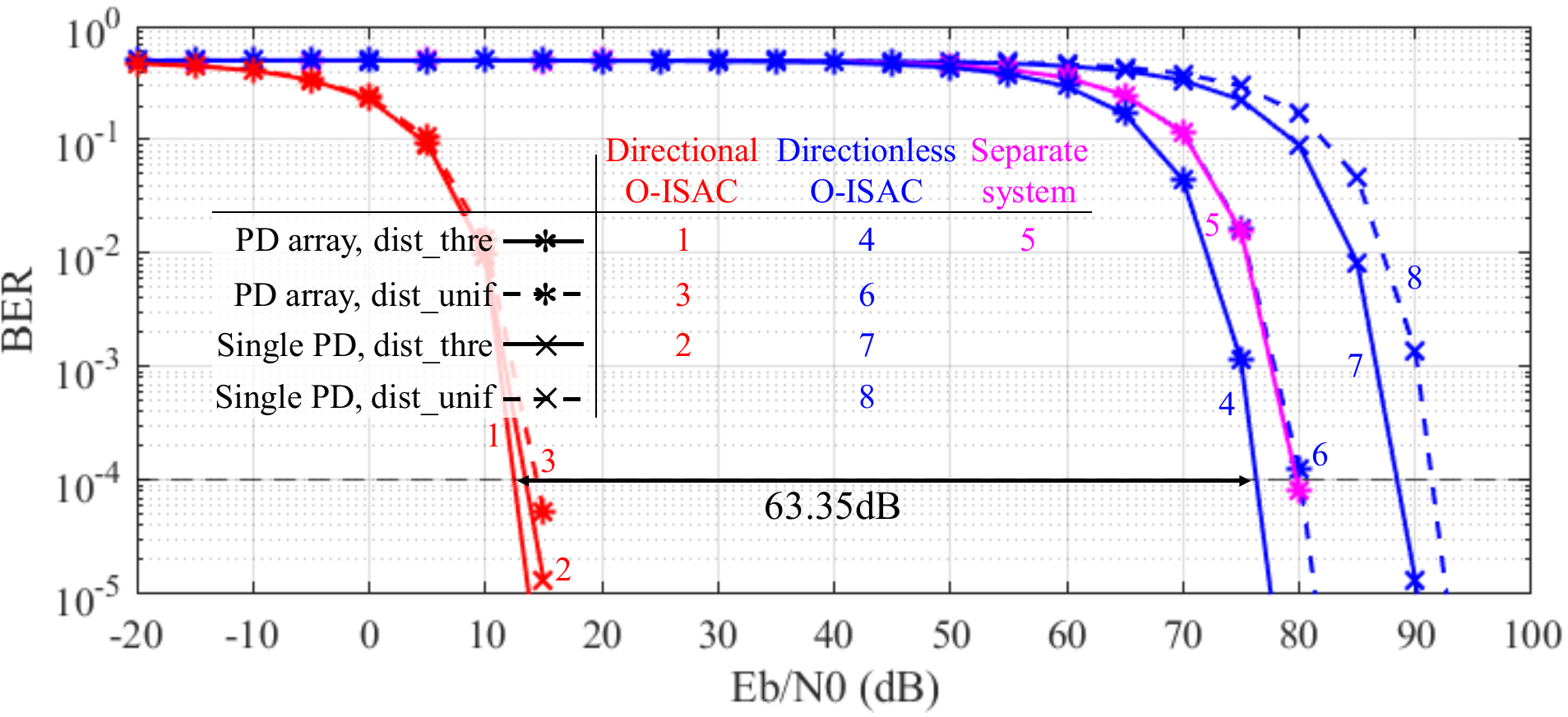} 
\caption{BER for optical communication under varied conditions.}
\label{f:oc}
\end{figure}

\begin{figure}[!t] 
\centering\includegraphics[width=.9\linewidth]{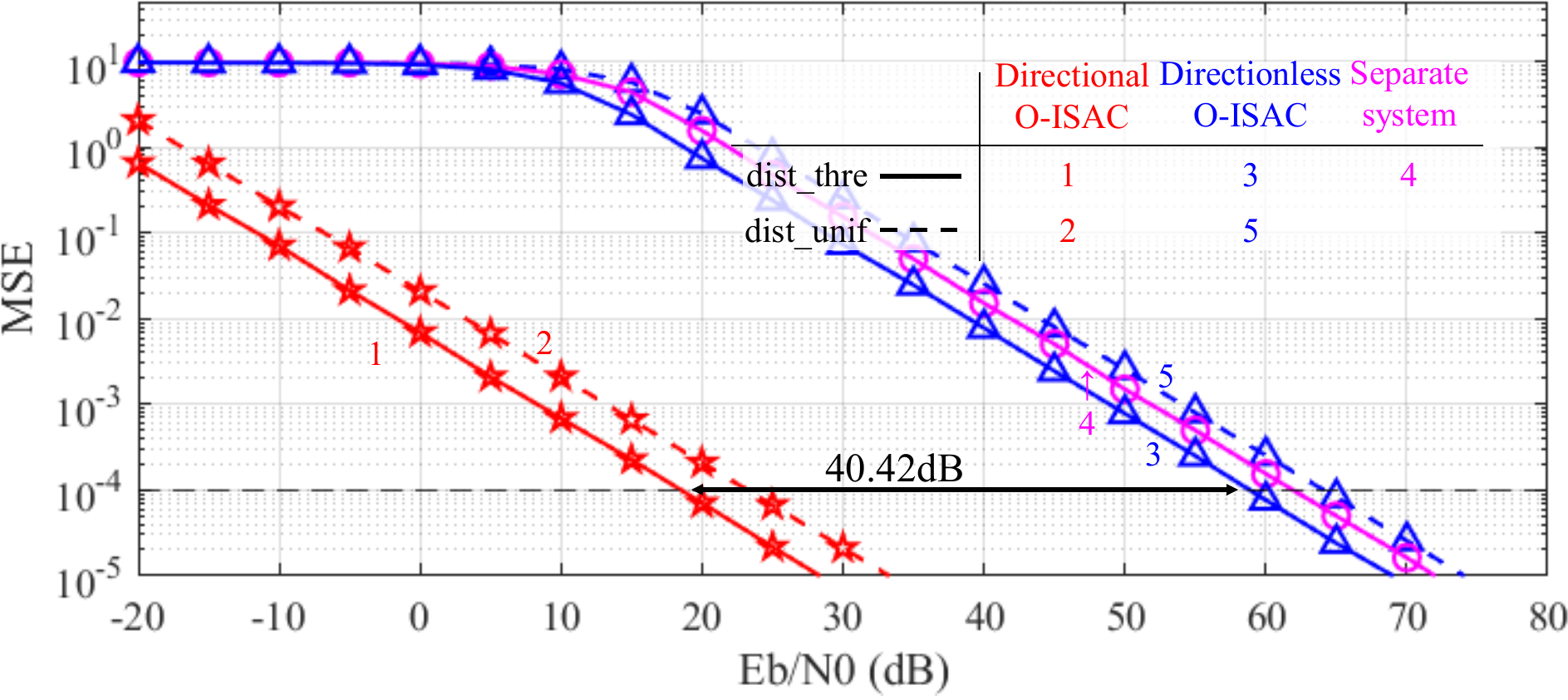} 
\caption{MSE of position estimation under varied conditions.}
\label{f:os}
\end{figure}

To quantitatively analyze Fig.~\ref{f:coverage}, we set a reference value for both communication BER and sensing MSE at $10^{-4}$ (note that the location estimation error is $1$cm when $\text{MSE}=10^{-4}$).
Base on Fig.~\ref{f:coverage}, Fig.~\ref{f:bar} summarizes the proportion of the room area that achieves better performance than the reference value under the two optimization criteria.
As can be seen, the proposed optimization goal in \eqref{e:optimization_h} yields better performance in terms of all three metrics. The gains are up to $7.26\%$, $33.79\%$, and $1.82\%$, respectively.
Overall, in the first phase, the BER and MSE performances of O-ISAC versus SNR are given in Figs. \ref{f:oc} and \ref{f:os}, respectively.
We will provide a more detailed comparison with the second phase of O-ISAC later.

\begin{figure}
    \centering
    \begin{subfigure}[t]{0.49\linewidth}
        \centering
        \includegraphics[width=\linewidth]{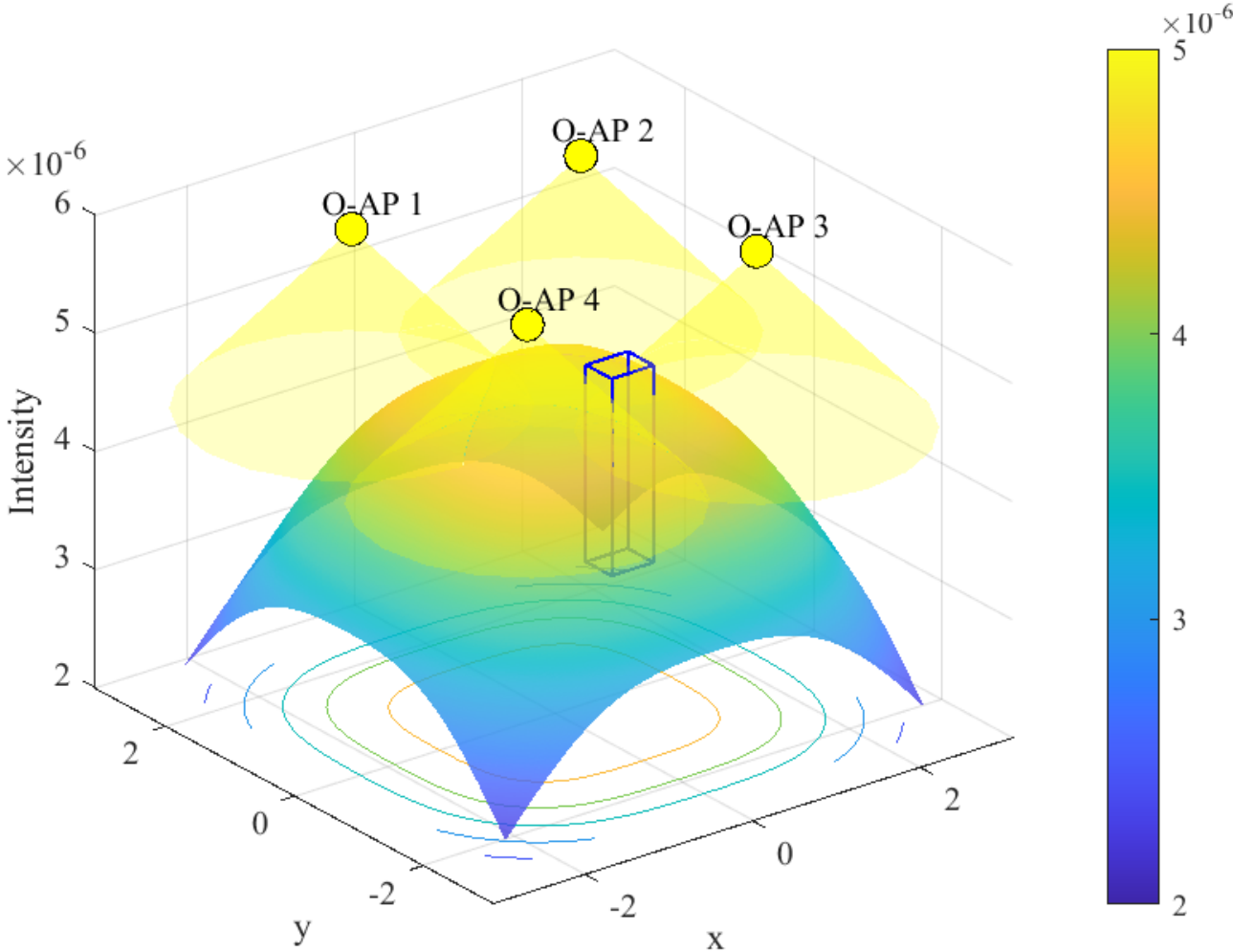}
        \caption{Phase 1}
    \end{subfigure}
    \begin{subfigure}[t]{0.49\linewidth}
        \centering
        \includegraphics[width=\linewidth]{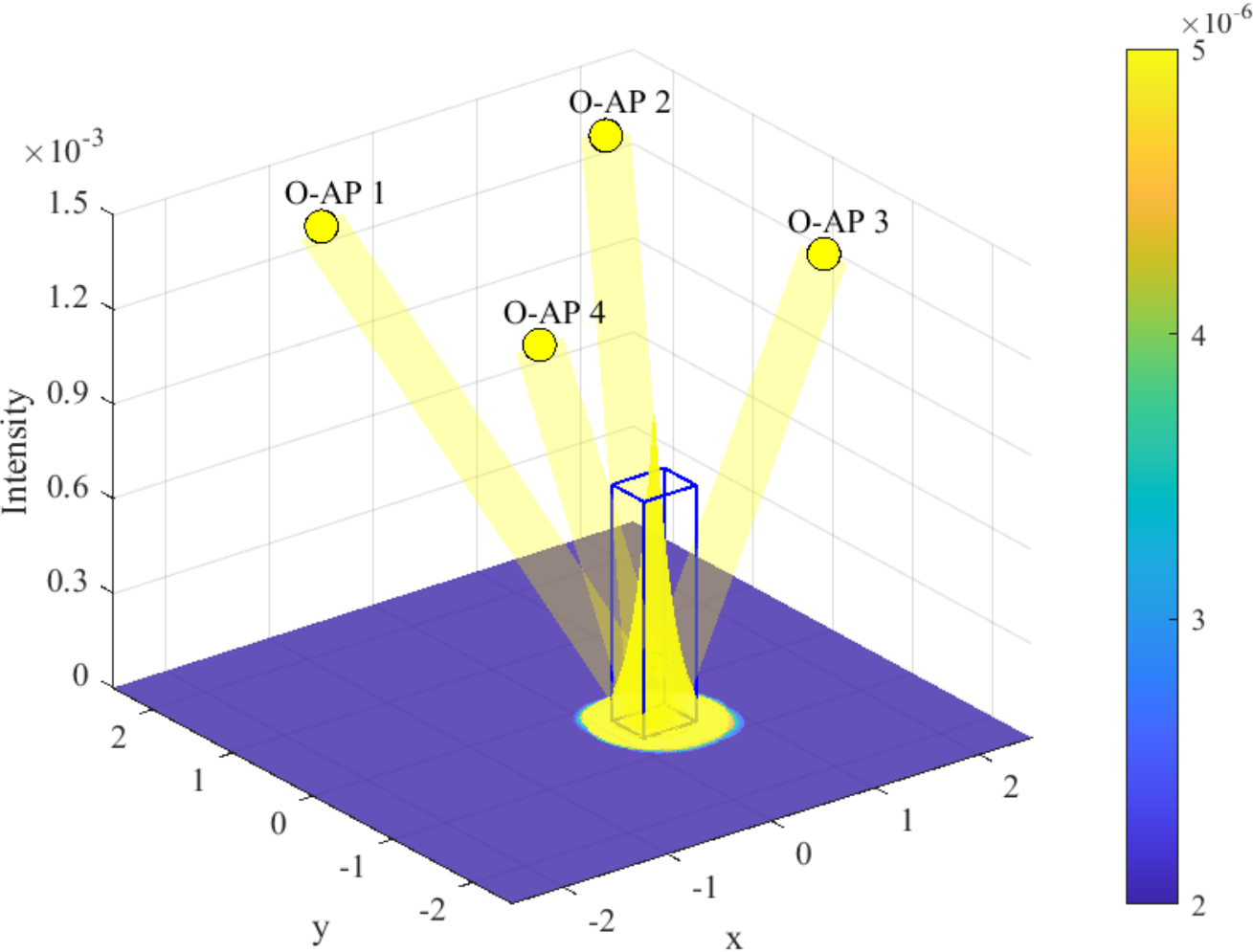}
        \caption{Phase 2}
    \end{subfigure}\\
    \caption{The distribution of light intensity in the environment, where the blue box represents the position of the target device.}
    \label{f:intensity}
\end{figure}

Compared to the first phase, the most significant feature of the second phase is the introduction of optical beamforming. This innovation concentrates the light emitted by O-APs onto the target device, thereby reducing the SNR required for attaining the target BER of communications and the MSE of optical sensing significantly.
Specifically, in the first phase, the light intensity that falls on the target only accounts for $0.96\%$ of the total light intensity, as illustrated in Fig.~\ref{f:intensity}(a). 
In the second phase, on the other hand, the received light intensity exhibits a distribution shown in Fig.~\ref{f:intensity}(b), where $66.41\%$ of the light intensity falls on the target.

Overall, this paper has considered several optimization schemes including 1) optimization of the light source distribution; 2) utilization of a PD array; 3) optical beamforming, also known as directional O-ISAC.
Thanks to these optimization schemes, the performances of both optical sensing and communications are improved.
Fig.~\ref{f:oc} presents the BER performance of the O-ISAC system.
The results reveal that, relative to the separated system, the non-directional O-ISAC achieves a 3.47 dB improvement, whereas the directional O-ISAC secures a substantial 63.35 dB gain. Implementing a PD array in the non-directional O-ISAC system yields approximately a 12 dB increase, with the strategic arrangement of light sources contributing an extra 3 dB gain. This arrangement primarily optimizes phase 1, indicating that the configuration of PDs and light sources exerts minimal influence on the performance of the directional O-ISAC system.

Considering that the variation of PDs does not affect the reflected light, the MSE estimated at the O-AP only considers the changes in the light source distribution, and Fig.~\ref{f:os} presents the MSE of position estimation in optical sensing.
Directionless O-ISAC outperforms the separate system by $3.14$ dB, while directional O-ISAC outperforms the directionless system by $40.42$ dB.
Moreover, the optimization of the light source distribution can provide a gain of 5 dB for both the Directionless O-ISAC and Directional O-ISAC systems.

\begin{figure}[!t] 
\centering\includegraphics[width=0.8\linewidth]{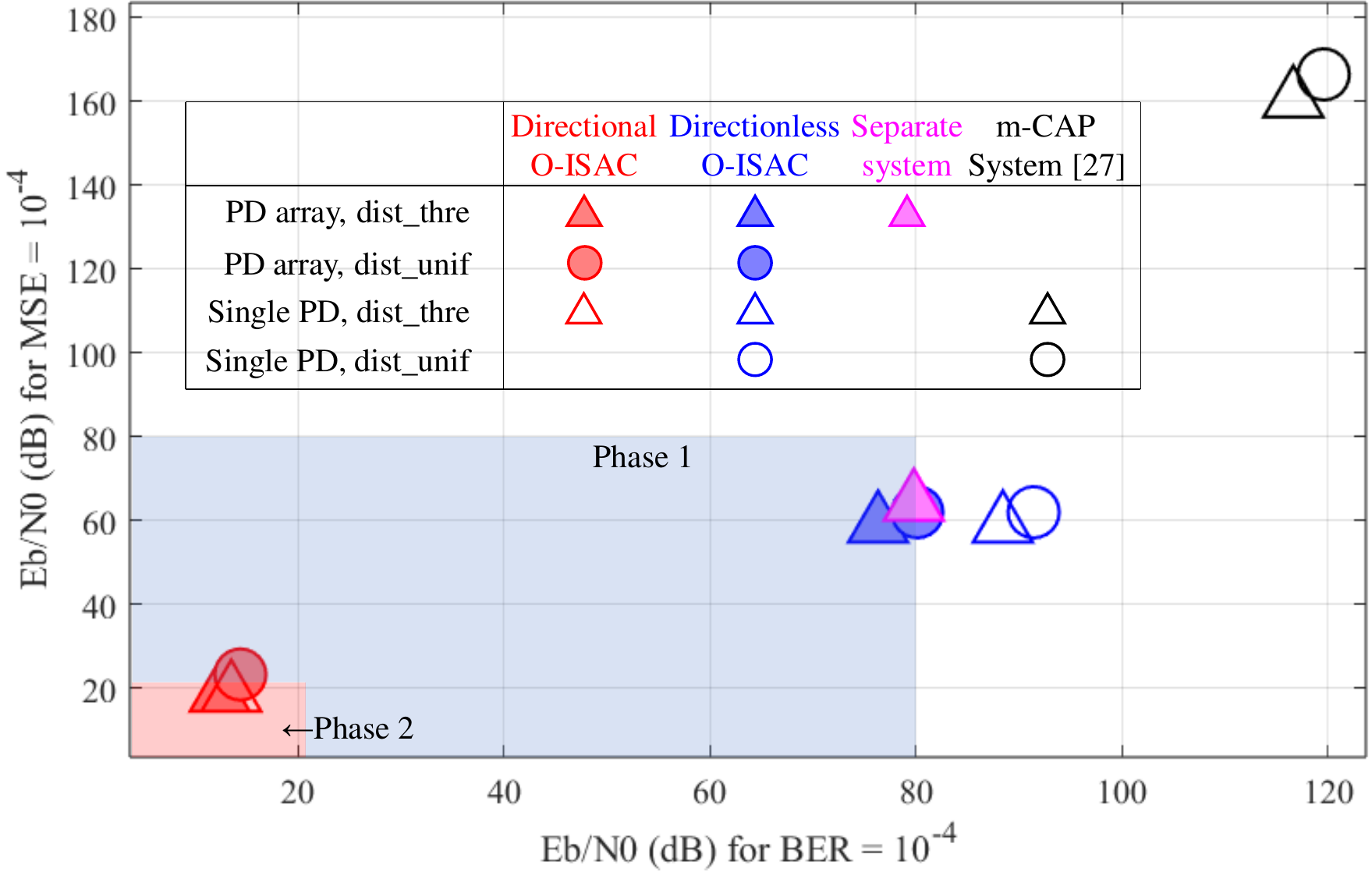} 
\caption{A comparison of the required Eb/N0 to achieve a BER of $10^{-4}$ and an MSE of $10^{-4}$ across different system configurations.}
\label{f:comparison}
\end{figure}

Overall, in our O-ISAC system, the O-APs periodically use a large power (e.g., $80$ dB in Fig. \ref{f:oc}) to broadcast the control information and sense the devices globally. The BER can be kept to $10^{-4}$ and the sensing MSE is lower than $10^{-4}$ (this corresponds to a localization accuracy of $1$ cm).
Then, in the second phase (which is much longer than the first phase), the O-APs use a relatively low power (e.g., $20$ dB in Fig. \ref{f:oc}) to serve the users and keep track of the users' locations.
The BER performance can be kept way below $10^{-4}$ and the sensing MSE is also kept below $10^{-4}$.
An illustration of the required power to achieve a BER of $10^{-4}$ and an MSE of $10^{-4}$ across different system configurations is provided in Fig.~\ref{f:comparison}, with a comparison to the m-CAP based joint Visible Light Sensing and Communication system from \cite{shi2022joint}. It can be observed that even the proposed directionless O-ISAC system outperforms the system in \cite{shi2022joint}, thanks to its pinhole based sensing system. Additionally, in phase 2, optical beamforming further reduces the required power by concentrating the energy.

\section{Conclusion}\label{sec:VI}
This paper presented and evaluated a new O-ISAC paradigm tailored for commercial LEDs. The primary goal of O-ISAC is to integrate optical communication and optical sensing capabilities within a single framework to improve both communication efficiency and device localization accuracy -- critical components in smart IoT ecosystems. Our approach employs a two-phase strategy, utilizing high-power transmission in the first phase for global device localization and control information dissemination, and transitioning to lower-power transmission in the second phase to serve users and maintain precise device tracking. Overall, our O-ISAC framework's ability to provide robust and efficient indoor optical communication while simultaneously enhancing device localization accuracy makes it a promising solution for IoT communication and sensing applications, such as smart homes and industrial IoT. Moving forward, we will focus on practical implementation and deployment considerations to further validate the system's real-world feasibility.

\begin{appendices}

\section{Proof of Lemma \ref{thm:2}}\label{Appendix_exit_angle}

This appendix proves Lemma \ref{thm:2}.
Considering an optical source and denoting the peak wavelength by $\lambda_0$, this appendix approximates the AoD $\varphi(\lambda|\phi)$ for $\forall \phi$ after passing through the lens.

As shown in Fig.~\ref{f:lens}(b), for different wavelengths $\lambda$, the incident angle $\beta$ w.r.t. the lens surface is the same. The emission angles $\theta(\lambda_0|\phi)$ and $\theta(\lambda|\phi)$ for wavelengths $\lambda_0$ and $\lambda$, respectively, follow
\begin{equation}\label{eqn9}
\begin{cases}
\sin\beta \cdot n(\lambda_0) &= \sin\big( \theta(\lambda_0|\phi) \big)\cdot1,  \\
\sin\beta \cdot (\lambda) &= \sin\big( \theta(\lambda|\phi) \big)\cdot1.
\end{cases}
\end{equation}

Since the exit ray of wavelength $\lambda_0$ is parallel to the $y$-axis, $\varphi(\lambda|\phi)=\theta(\lambda_0|\phi)-\theta(\lambda|\phi)$ is the angle between the $y$-axis and the outgoing light ray of wavelength $\lambda$. It can be written as
\begin{eqnarray}\label{e:theta_out_approx}
&&\hspace{-0.4cm} \varphi(\lambda|\phi)= \theta(\lambda_0|\phi)  - \theta(\lambda|\phi) \nonumber \\
&&\hspace{-0.75cm} = \arcsin\big(  n(\lambda_0) \sin\beta \big)  -  \arcsin\big(  n(\lambda) \sin\beta  \big) \nonumber \\
&&\hspace{-0.8cm} \overset{(a)}{=} \arcsin\big(n(\lambda_0)\sin\beta\big) - \arcsin\left(n(\lambda_0)\frac{\lambda_0}{\lambda }\sin\beta\right)  \nonumber \\
&&\hspace{-0.8cm} \overset{(b)}{\approx} n(\lambda_0)\sin\beta + \frac{1}{6} \big(n(\lambda_0)\sin\beta \big)^3 + \frac{3}{40} \big(n(\lambda_0)\sin\beta \big)^5 \nonumber \\
&&\hspace{0cm} - \biggl(n(\lambda_0)\frac{\lambda_0}{\lambda } \sin\beta\biggr) - \frac{1}{6} \left(n(\lambda_0)\frac{\lambda_0}{\lambda }\sin\beta\right)^3  \nonumber \\
&&\hspace{0cm} -\frac{3}{40} \left( n(\lambda_0) \frac{ \lambda_0} { \lambda } \sin\beta \right)^5 \nonumber \\
&&\hspace{-0.75cm}=n(\lambda_0)\left[1-\frac{\lambda_0}{\lambda }\right]\sin\beta+\frac{1}{6} \big(n(\lambda_0) \big)^3\left[1-\frac{\lambda_0^3}{\lambda^3}\right]\sin^3\beta \notag \nonumber \\
&&\hspace{-0.4cm} +\frac{3}{40} \big(n(\lambda_0) \big)^5 \left[1-\frac{\lambda_0^5}{\lambda^5}\right]\sin^5\beta \nonumber \\
&&\hspace{-0.75cm} \approx n(\lambda_0)\left[1-\frac{\lambda_0}{\lambda}\right]\beta+\frac{1}{6} \big(n(\lambda_0) \big)^3 \left[1-\frac{\lambda_0}{\lambda }\right]^3\beta^3 \nonumber \\
&&\hspace{0cm}  +\frac{3}{40} \big(n(\lambda_0) \big)^3 \left[1-\frac{\lambda_0}{\lambda }\right]^5\beta^5  \\
&&\hspace{-0.8cm}\overset{(c)}{\approx} \arcsin\left(n(\lambda_0)\left(1-\frac{\lambda_0}{\lambda }\right)\sin\beta\right) 
\approx n(\lambda_0)\left(1-\frac{\lambda_0}{\lambda }\right)\beta,\nonumber
\end{eqnarray}
where 
(a) follows from \eqref{eqn9};
both (b) and (c) follow from Taylor series expansion of arcsine.

Next, we derive the incident angle $\beta$ on the lens' surface.
For a given wavelength, there is a one-to-one correspondence among the AoE $\phi$ of the light source, the incident angle $\beta$ of the lens' surface, the normal vector $\overrightarrow{n}_f$, and the outgoing angle $\theta(\lambda_0|\phi)$.
Therefore,
\begin{eqnarray}\label{e:beta}
&&\hspace{-0.9cm}\beta \overset{(a)}{=} \arccos \left[ \frac{ n(\lambda_0) \sin \phi \sin \phi + ( n(\lambda_0) \cos \phi - 1 ) \cos \phi } { \sqrt{ \big(n(\lambda_0)\big)^2 \sin^2 \phi + ( n(\lambda_0)  \cos \phi - 1 ) ^2 } } \right] \nonumber \\
&&\hspace{-0.85cm} = \arccos\left[\frac{n(\lambda_0)-\cos\phi}{\sqrt{\big(n(\lambda_0)\big)^2-2n(\lambda_0)\cos\phi+1}}\right] \nonumber \\
&&\hspace{-0.9cm} \overset{(b)}{\approx} \sqrt{2\left(1-\frac{n(\lambda_0)-\cos\phi}{\sqrt{\big(n(\lambda_0)\big)^2-2n(\lambda_0)\cos\phi+1}}\right)} \nonumber \\
&&\hspace{-0.3cm}+\frac{\sqrt{2}}{12}\left[1-\frac{n(\lambda_0)-\cos\phi}{\sqrt{n^2-2n\cos\phi+1}}\right]^{\frac{3}{2}} \nonumber \\
&&\hspace{-0.3cm}  +\frac{3\sqrt{2}}{160}\left[1-\frac{n(\lambda_0)-\cos\phi}{\sqrt{\big(n(\lambda_0)\big)^2-2n(\lambda_0)\cos\phi+1}}\right]^{\frac{5}{2}}\nonumber \\
&&\hspace{-0.85cm} \approx \sqrt{2\left(1-\frac{n(\lambda_0)-\cos\phi}{\sqrt{\big(n(\lambda_0)\big)^2-2n(\lambda_0)\cos\phi+1}}\right)} \nonumber \\
&&\hspace{-0.9cm} \overset{(c)}{\approx} \sqrt{2\!\!\left(\!1\!-\frac{n(\lambda_0)-\left(1-\frac{1}{2}\phi^2+\frac{1}{24}\phi^4\right)}{\sqrt{\big(n(\lambda_0)\big)^2\!\!-2n(\lambda_0)\left(1-\frac{1}{2}\phi^2+\frac{1}{24}\phi^4\right)+1}}\right)} \nonumber \\
&&\hspace{-0.85cm} \approx \sqrt{2\left(1-\sqrt{1-\frac{\phi^2\left(1-\frac{1}{4}\phi^2\right)}{(n(\lambda_0)-1)^2+n(\lambda_0)\phi^2}}\right)} \nonumber \\
&&\hspace{-0.85cm} \approx \sqrt{2\left(1-\sqrt{1-\frac{\phi^2}{(n(\lambda_0)-1)^2+n(\lambda_0)\phi^2}}\right)} \nonumber \\
&&\hspace{-0.9cm} \overset{(d)}{\approx} \sqrt{ \frac{ \phi^2} {(n(\lambda_0)-1)^2}},
\end{eqnarray}
where (a) follows from $\cos \beta = \frac{\overrightarrow{n}_f \cdot \overrightarrow{\text{OA}}}{\left| \overrightarrow{n}_f \right| \left| \overrightarrow{\text{OA}} \right|}$;
(b) and (c) follow from Taylor series expansion;
(d) follows from 
\begin{eqnarray}
    &&\hspace{-1cm}\sqrt{2\left(1-\sqrt{1-\frac{\phi^2}{\big(n(\lambda_0)-1\big)^2+n(\lambda_0)\phi^2}}\right)} \nonumber \\
    &&\hspace{0.5cm} = \sqrt{ \frac{ \phi^2} {\big(n(\lambda_0)-1\big)^2}} + \frac{(1-4n(\lambda_0))|\phi|^3}{8\big(n(\lambda_0)-1\big)^3} + O(\phi)\nonumber .
\end{eqnarray}

Substituting \eqref{e:beta} into \eqref{e:theta_out_approx} yields
\begin{eqnarray}\label{e:approx}
    &&\hspace{-0.5cm}\varphi(\lambda|\phi) \approx n(\lambda_0)\left(1-\frac{\lambda_0}{\lambda }\right)\sqrt{\frac{\phi^2}{\big(n(\lambda_0)-1\big)^2}}\nonumber\\
    &&\hspace{0.625cm} = \frac{n(\lambda_0)(\lambda -\lambda_0)}{\big(n(\lambda_0)-1\big)\lambda }\phi,
\end{eqnarray}
proving Lemma \ref{thm:2}.

\end{appendices}

\bibliographystyle{IEEEtran}
\bibliography{ref.bib}

\vfill

\end{document}